"Cutting through the jargon and technicalities of quantum science, Chris Ferrie uses humor and wit to educate the reader on the current state of the quantum technological industry. This book helps debunk some of the biggest myths surrounding quantum computing."

**KENNA HUGHES-CASTLEBERRY,**
author of *On The Shoulders of Giants*

# What You SHOULDN'T Know About QUANTUM COMPUTERS

# Chris Ferrie

# Also by Chris Ferrie

Great Quotes from Great Scientists: Quotes, Lessons, and Universal Truths from the World's Greatest Scientific Minds

Quantum Bullsh*t: How to Ruin Your Life With Advice From Quantum Physics

42 Reasons To Hate The Universe (And One Reason Not To)

Where Did The Universe Come From? And Other Cosmic Questions

# What You Shouldn't Know About Quantum Computers

CHRIS FERRIE

# Contents

# Foreword

For almost twenty years, my blog, "Shtetl-Optimized," has been (I'd like to think) the central clearinghouse on the Internet for puncturing inflated claims about how quantum computers would work and what kinds of problems they would help with. That's sort of my beat. I never planned it that way, but the more hype flooded the tech press, the less choice I felt.

Yet in all that time I never thought to write, much less did I actually write, a pithy book called "What You Shouldn't Know About Quantum Computers." My colleague Chris Ferrie did. He's the same guy who coauthored the surprise bestseller "Quantum Computing for Babies." Now he's back, with something for those babies to read when they're slightly older.

In this short volume, Chris provides a heaping helping of truth on every page. He explains that, at least on a mathematical level, quantum mechanics is not an endless mystery: it's just a specific, fixed thing that's different from what you're used to, until you do get used to it. It's not anything-goes, and it's far too simplistic to say it involves "everything happening at once." Rather, the key is that the state of any isolated physical system (like an electron or a molecule, or even in principle the whole universe) is really a gargantuan list of numbers—complex numbers, called amplitudes. There's one amplitude for each thing you could possibly find the system doing when you look. These amplitudes are closely related to the probabilities of finding

the system doing the various things, but they're not probabilities. (For starters, they can be negative.)

Once he's explained that, Chris is able to make quick work of one journalistic trope about quantum computing after another. For example, would quantum computers work by trying all possible answers in parallel? Sorry, no, that's too good to be true: Quantum computers work by choreographing a pattern of interference, where the contributions to the amplitude of each wrong answer cancel each other out, while the contributions to the right answer's amplitude reinforce each other. Only for special problems, as it turns out, do we know how to choreograph such an interference pattern to deliver a huge speedup over the best known classical algorithms. This, in turn, is why we don't expect quantum computers ever to replace classical computers, but "merely" to complement them, accelerating specific tasks like quantum simulation and codebreaking.

Likewise, could quantum computers use entanglement to send messages faster than the speed of light, in what Einstein called "spooky action at a distance"? No, that's another misconception. What's true is "merely" that a classical simulation of quantum mechanics would require faster-than-light communication. Once again, quantum mechanics stakes out a subtle intermediate position between "familiar classical world" and "anything-goes fantasyland": a position that no science fiction writer would've had the imagination to invent. Then there's the myth that quantum computers are impossible, for some simple reason that all the world's physicists have overlooked. While this might seem like the opposite of "quantum computers can do anything" myth, I actually see them as two sides of the same coin: the more you understand about the limitations of quantum computers, the less shocking it is that they can exist. Or maybe quantum computers can't exist, at the scale and reliability that we theorists envision. But if so, then that itself would be a revolutionary discovery in physics, vastly more surprising than mere "success" in building QCs! The huge experimental efforts now underway strongly suggest that one way or the other, we're going to find out.

Anyway, Chris explains all this in a way that I think readers will find accessible and fun—maybe even those who tell me they don't understand a word of my blog.

I have quibbles with this book, but they're merely the quibbles I'd have with a colleague, not the forehead-banging frustrations I have with popular manglings of quantum computing. Chris seems to be optimistic about quantum algorithms for finance; I'm much less so. Chris hates the many-worlds interpretation of quantum mechanics, while I'm (as Chris says in the book) about as positive as you can be about many-worlds without actually believing it. (On the other hand, Chris and I stand shoulder to shoulder against MWI-inspired misconceptions about quantum computing, like "QCs just try all the answers in parallel" or "a thousand qubits are equivalent to $2^{1000}$ classical bits.")

On reflection, though, I'm glad that I can find things to quibble with in this book. Otherwise I'd be overpowered with regret that I hadn't written it! And that, of course, is just about the highest praise I can give.

Scott Aaronson

Austin, Texas, 2024

# The Quantum Age

> *The nineteenth century was known as the machine age, the twentieth century will go down in history as the information age. I believe the twenty-first century will be the quantum age.*
>
> —Paul Davies

Computers are everywhere—you might even be reading this on one! But what is it that computers *do*? Let's step back and ask instead what it is we want them to do.

## Algorithms and programs

Consider the fact that there is nothing your digital computer can compute that a room full of people with pen and paper could not. Remember long division? Imagine those people are doing long division—a sequence of memorized steps on new pieces of data. The steps your teacher drilled into you are called an algorithm. The steps are abstract rules that people with pen and paper could carry out, people with an abacus, electronic circuits with a Windows operating system, and so on. Given a particular set of tools, the specific instructions needed to carry out the steps using those tools are called the program. (You have several programs unique to your brand of computer that are carrying out the steps of algorithms right now.)

Regardless of the tools used, if the algorithm is followed faithfully, the answer will always be the same. People with pen and paper are prone to errors and are slow. Electronic circuits are near perfect and extremely fast. The choice is clear. Let's put it another way. There are plenty of algorithms we want—or need—to carry out. Wouldn't it be great if we could engineer a device to do it for us quickly and accurately? Yes, of course it would.

Remarkably, every algorithm we can think of can be cast as a program that changes sequences of 1s and 0s one step at a time. If we could build a machine that can represent any sequence of 1s and 0s and accept instructions (programs) for changing one sequence into another, we could use it to carry out the steps in any algorithm instead of doing it ourselves. Indeed, this is what a modern digital computer is—a machine that can perform any computation for us.

Although a digital computer can carry out the steps of any algorithm, some algorithms have digital programs that require too many steps to be feasible. For example, solving certain equations using digital computers is completely infeasible, even if the computer could do it in principle. We know, roughly speaking, the equations we would need to solve to determine what would happen when a hypothetical drug molecule interacted with a biological molecule. There are even algorithms to solve these equations. Many of the world's supercomputers chug away, running inefficient programs to solve such equations. But it's just too hard—so we "solve" the equations with genetically modified mice instead.

## The quantum ask

As before, we ask the following: wouldn't it be great if a machine could be built especially for the purpose of carrying out these equation-solving algorithms? Yes, that would be fantastic. But what would it need to do exactly? As it turns out, scientists have found algorithms whose steps change not sequences of 1s and 0s but sequences of complex numbers—numbers like the square root of -1 and π. These algorithms, though constrained not to be all-powerful, have far fewer steps than the original algorithms.

In some sense, these algorithms are not special—children could be trained to carry out the steps like long division. But that would be as inefficient as it is cruel. So, the original question becomes more precise. Can we build a machine that can represent any sequence of complex numbers and accept programs to change one sequence into another? Indeed, we can—and that device is called a *quantum computer*.

Nowadays, many algorithms are phrased as steps that change sequences of complex numbers rather than 1s and 0s. These so-called *quantum* algorithms could be performed by hand or using a digital computer, but it would take a long time. A quantum computer is a special-purpose device used to carry out these steps directly and efficiently.

Many things can be used to represent a 1 or 0. Digital computing technology has settled on the transistor, billions of which are in your computer now. But how do you "naturally" represent complex numbers? As it turns out, there's a theory for that! Quantum physics demonstrates that complex numbers are encoded by the fine details of light and matter. Really, they are all around us—we just don't have the finesse to get at them with our clumsy hands. But science and engineering have progressed through the past 100 years to give us the required level of control. We can now build devices to encode complex numbers and change them.

At the time of writing, all quantum computers in existence are prototypes—they are small and not reliable enough to be practical. They also don't have quantum programs at the level of abstraction that we have for digital computers. There is no "quantum" version of the word processing program I'm writing this on or the web browser program you potentially downloaded it on. The most interesting thing to imagine is what those quantum programs might look like—after all, when digital algorithms were being invented and the early computing machines were built to implement them, no one could have imagined that programs for future versions of those machines would exist to write blogs, share videos, perform bank transactions, and all the other things we take for granted. We know quantum computers will be able to solve some important problems for us, but what we will eventually have them do for us is unimaginable today because human foresight rather poor.

## Swiss Army computer

Even if I had a use for every one of the tools on a Swiss Army knife, I still would not buy one—it's too bulky, and I'm not headed into the woods anytime soon. Such a knife illustrates that specialized tasks are best carried out using specialized tools. If you open a hundred bottles of wine per day, for example, a Swiss Army knife has a corkscrew—but you are far better off buying a machine optimized for opening bottles of wine.

A computer is like a Swiss Army knife but for calculations. A computer can solve all sorts of mathematical problems. And that's extremely useful because many everyday problems can be phrased as math problems. Obvious examples are determining how much tip to leave at a restaurant, figuring out what time to catch the bus to arrive early for that meeting, adding up the values in a spreadsheet, and so on. Less obvious examples that are really just hidden math problems are recognizing faces in a digital photo, formatting words in a document, and seamlessly showing two people's faces to each other on other sides of the world in real time.

The central processing unit, or CPU, inside your tablet, smartphone, or laptop is tasked with carrying out any possible set of instructions thrown at it. But, because it can do anything, it's not the best at doing specific things. This is where the other PUs come in. Probably the most famous is the GPU, or graphics processing unit.

Maybe graphics aren't something you think about a lot. But even displaying something as simple as text on your screen requires coordinating the brightness and color of millions of pixels. That's not an easy calculation for a CPU. So, GPUs were made as special-purpose electronic devices that do the calculations required to display images well and not much else. The CPU outsources those difficult calculations to the GPU, and video gamers rejoice!

## The QPU

Calculations involving the multiplication and addition of complex numbers are very time-consuming for a CPU. As you now know, these kinds of calculations are essential for solving problems in quantum physics, including simulating chemical reactions and other microscopic phenomena. It would be convenient for these kinds of calculations if a quantum processing unit (QPU) were available. These are confusingly called quantum computers, even though they are chips sent very specific calculations by a CPU. I predict they will eventually be called just QPUs.

You won't find a QPU inside your computer today. This technology is being developed by many companies and academic researchers around the world. The prototypes that exist today require a lot of supporting technology, such as refrigerators cooled using liquid helium. So, while QPUs are "small," the pictures you will see of them show large laboratory equipment surrounding them.

What will the future QPU in your computer do? First of all, we could not have guessed even ten years ago what we'd be doing today with the supercomputers we all carry around in our pockets. (Mostly, we are applying digital filters to pictures of ourselves, as it turns out.) So, we probably can't even conceive of what QPUs will be used for ten years from now. However, we do have some clues as to industrial and scientific applications.

At the Quantum Algorithm Zoo ([quantumalgorithmzoo.org](quantumalgorithmzoo.org)), 65 problems (and counting) are currently listed that a QPU could solve more efficiently than a CPU alone. Admittedly, those problems are abstract, but so are the detailed calculations that any processor carries out. The trick is in translating real-world problems into the math problems we know a QPU could be useful for. Not much effort has been put into this challenge simply because QPU didn't exist until recently, so the incentive wasn't there. However, as QPUs start to come online, new applications will come swiftly.

## Simulate all the things

My favorite and inevitable application of QPUs is the simulation of physics. Physics simulations are ubiquitous. Gamers will know this well. When you think of video games, you should think of virtual worlds. These worlds have physical laws, and the motion of the objects and characters in the world needs to be calculated—this is a simulation. Physics must be simulated when designing aircraft, bridges, and any other engineered system. Physics is simulated in science, too—entire galaxies have been simulated to understand their formation. But quantum physics has resisted simulation because CPUs are really bad at it.

Once we can simulate quantum physics on QPUs, we can simulate chemical interactions to rapidly design new materials and medicines. We might also be able to simulate the physics at the creation of the universe or the center of a black hole, and who knows what we will find there.

## The takeaway

Qubits, superposition, entanglement, parallelism, and other quantum magic you've read about elsewhere are not useful concepts to think about if you only want a five-minute summary of quantum computers. The basic thing you need to know about QPUs is the same thing you know about GPUs—they are special-purpose calculators that are good at solving a particular kind of mathematical problem.

If, at some point, you end up with a job title that has the word quantum in it, it will probably be a software job (much like there are 20 software engineers for every computer hardware engineer today). The most challenging problem a Quantum Solutions Engineer might face is in translating the calculations their business currently performs into problems that can be outsourced to a QPU. They may not even use or need to understand concepts like superposition and entanglement! So, you have a pass if you'd like to be spared the details.

The bigger problem today is that the details of quantum concepts are not avoided, but replaced with poor analogies and mysticism. The Quantum Age looms, but today, we are in the Quantum Dark Ages, where myth and superstition run rampant.

# Whence Quantum Computing?

> *"History is not the past but a map of the past, drawn from a particular point of view, to be useful to the modern traveller."*
>
> — Henry Glassie

Almost any discussion of quantum computing—whether in a research article, popular science magazine, or business journal—invokes some comparison to what is called "classical" computing. It's nearly impossible to talk about quantum computing without using the phrase, so we better start there.

The term *classical* in classical computing is borrowed from conventions in physics. In physics, we often denote pre-1900 physics as "classical" and post-1900 physics as "modern." Modern physics includes general relativity and quantum physics. General relativity is Einstein's theory of curved space and time, which explains the force of gravity. This theory, while instrumental in enabling us to comprehend awe-inspiring images of galaxies, has its most direct technological application in GPS, the Global Positioning

System crucial for satellite navigation. Yet, even this remarkable application is not standalone—it also requires quantum technology for accurate functioning.

Quantum computing theory might have some things to say about places of extreme gravity, like black holes, but we will stay grounded on Earth and won't say more about general relativity in this book. The reason is that, in contrast to general relativity, quantum physics has a broader range of applications that span diverse fields. It is the backbone of lasers and light bulbs, the lifeblood of medical scanners and radiation therapy, and the cornerstone of semiconductors and electron microscopes. It governs the precision of atomic clocks and the power of atomic bombs, among many others. And it is the fuel for quantum computers.

The potency of quantum physics and its technological implications stem from exploring a new world—the microscopic world of atoms. By uniting three of the four fundamental forces of nature, quantum physics presents us with rules governing the basic constituents of matter and energy. The fourth force, gravity, while undeniably significant, is relatively weak—evident when we effortlessly overcome Earth's gravitational pull each time we stand up. The paradigm shift from classical to quantum physics propelled humanity from the era of steam engines and machinery to the digital, information-driven age. Now, it is driving us into the quantum age.

In the remainder of this chapter, I want to give you a bit of context about how and why quantum computing came to be. Until recently, quantum and computing were mostly separate fields of study, although they overlap significantly in their engineering applications. So, we'll go over the "quantum" and "computing" parts first before bringing them together.

## The "quantum" in quantum computing

The story of quantum physics started in 1900 with Max Planck's "quantum hypothesis." Planck introduced the idea that energy is quantized, meaning it comes in discrete packets, which came to be called quanta. The problem Planck was studying was dubbed the *black body radiation problem.* A black body is an idealization of an object that absorbs and emits radiation. It's a pretty good approximation for hot things in thermal equilibrium, like the Sun or a glowing iron pot, which give off roughly the same spectrum (colors) when at the same temperature. In other words, if you heat something up so that it glows "white hot," it will be the same temperature as the Sun (roughly 6000 °C). The problem was that no one could figure out why using the physics of the time (now called "classical" physics).

As Planck was doing his calculations, he discovered that if energy has some smallest unit, the formulas worked out perfectly. This was a wild guess he made in desperation, not something that could have been intuited or inferred. But it deviated from classical physics at the most fundamental level, seemingly rendering centuries of development useless, so even Planck himself didn't take it seriously at first. Indeed, it took many years before others began to pay attention and many more after that before the consequences were fully appreciated. Perhaps not surprisingly, one of those who did grasp the importance early on was Albert Einstein.

Among his illustrious list of contributions to science, Einstein used Planck's quantum hypothesis to solve the problem of the *photoelectric effect*. The photoelectric effect is a phenomenon in which electrons are emitted from a material when exposed to light of a certain frequency. Classical physics could not explain why the energy of these ejected electrons depended on the frequency (color) of the light rather than its intensity (brightness) or why there was a cut-off frequency below which no electrons were emitted, regardless of the intensity.

Einstein proposed that light itself is quantized into packets of energy, which later came to be called *photons*. Each photon has energy proportional to its frequency, in accordance with Planck's quantum hypothesis. When a photon hits an electron, if its energy (determined by its frequency) is sufficient, it can overcome the energy binding the electron to the material and eject it. Quantization had taken hold. This marked a profound transition in physics, which is still the source of confusion and debate today. Before photons, light was argued to be *either* a wave or a particle. Now, it seems that it is both—or neither. This came to be known as *wave-particle duality*.

Until this point, quantum theory seemed to apply only to light. However, in parallel with these developments, experiments were starting to probe the internal structure of atoms in evermore detail. One curiosity was *spectral lines*. Specific lines or gaps were present when observing the light that the purest forms of elements (hydrogen, helium, lithium, and so on) emitted or absorbed. Many formulas existed, but none could be derived from classical physics or any mechanism until Niels Bohr paid a visit to England in 1911 to discuss the matter with the leading atomic physicists of the day.

Bohr's model proposed that electrons in an atom could only occupy certain discrete energy levels, consistent with Planck's quantum hypothesis. When an electron jumps from a higher energy level to a lower one, it emits a photon of light with a frequency that corresponds to the difference in energy between the two levels. This model provided a way to calculate the energies of these spectral lines and brought the atom into the realm of quantum physics. He presented the model formally in 1913.

A decade later, Louis de Broglie proposed in his Ph.D. thesis that not just light but all particles, including electrons, have both particle and wave characteristics. This was a revolutionary idea that was quickly put to the test and verified. At this point, it was clear that quantum theory was more than a technique that could be applied as corrections when classical physics didn't fit the data. Scientists started to theorize abstractly using concepts of quantum physics rather than turning to it only as a last resort.

In 1925, Werner Heisenberg invented *matrix mechanics* to deal with the calculations underpinning the theory. Using it, he showed that it is impossible to measure the position and momentum of a particle simultaneously with perfect accuracy. This *uncertainty principle* became a fundamental and notorious aspect of quantum theory. At the same time, Erwin Schrödinger developed a mathematical equation to describe how de Broglie's waves might change in time. The variable in the equation—called the *wave function*—describes the probability of finding a particle in a given location or in a particular state.

In contrast to matrix mechanics, Schrödinger's picture was called *wave mechanics*. There was a brief debate about which of the two alternatives was correct. However, Paul Dirac showed that both are equivalent by axiomatizing the theory, demonstrating that a simple set of principles can derive all that was known about it. He also combined quantum mechanics with special relativity, leading to the discovery of antimatter and paving the way for further developments in quantum field theory, which ultimately led to the so-called *Standard Model of Particle Physics* that supported the seemingly unending stream of particle discoveries in high-energy particle accelerator experiments.

However, by the middle of the 20th century, quantum physics was more or less a settled science. The applications of the theory far outstripped the discoveries, and the majority of the people using it resembled engineers more so than they did physicists. We'll get to those applications briefly, but first, we need to jump ahead before jumping back in time. During the 1960s and '70s, engineers in the laboratories of information technology companies like Bell and IBM began to worry about the limits of the new communications and computing devices being built. These things require energy to function, and energy is the primary concern of physics. Did the laws of physics have anything to say about this? And, if they did, wouldn't the most fundamental laws (quantum laws) have the most to say? Indeed, they did, and this is where the two theoretical fields of physics and computing converged. But to appreciate it, we must tell the other half of the quantum computing backstory.

## The "computing" in quantum computing

Often, in quantum physics, it is useful to compare the results of some theory or experiment to what classical physics might predict. So, there is a long tradition in the *quantum vs. classical* comparison. This tradition was adopted first not by computer scientists or engineers but by quantum physicists who started to study quantum computing in the '80s. Unlike in physics, the adjective *classical* in classical computation does not mean it is pre-modern—it is just a way to distinguish it from quantum computation. In other words, whatever quantum computing was going to be compared to, it was bound to be referred to as "classical."

The device I am using to create this is anything but classical in the usual sense of the term. A modern laptop computer is a marvel of engineering, and digital computing would not be possible without quantum engineering. The components inside your computing devices, like your smartphone, are small enough that quantum mechanics plays an important role in how they were conceived, engineered, and function. What makes my laptop "classical" is not the hardware but the software. What the device does at the most abstract level—that is, *compute*—can be thought of in classical physics terms. Indeed, anything your smartphone can do, a large enough system of levers and pulleys can do. It's just that your computer can do it much faster and more reliably. A quantum computer, on the other hand, will function differently at both the device level *and* at the abstract level.

Abstract computational theory, now referred to as theoretical computer science, was born in the 1930s. British World War II codebreaker Alan Turing devised a theoretical model of a computer now known as a *Turing machine*. This simple, abstract machine was intended to encapsulate the generic concept of computation. He considered a machine that operates on an infinitely long tape, reading, writing, or moving based on a set of predetermined rules. Remarkably, with this simple model, Turing proved that certain problems couldn't be solved computationally at all.

Together, Turing and his doctoral supervisor, Alonzo Church, arrived at the *Church-Turing thesis*, which states that everything computable can be computed by a Turing machine. Essentially, anything that computes can be simulated by Turing's theoretical device. (Imagine a modern device that emulates within it a much older device, like a video game console, and you have the right picture.) A more modern version relating to physics states that all *physical processes* can be simulated by a Turing machine.

In 1945, John von Neumann expanded on Turing's work and proposed the architecture that most computers follow today. Known as the von Neumann architecture, this structure divides the computer into a central processing unit (CPU), memory, input devices, and output devices. Around the same time, Claude Shannon was thinking about the transmission of information. In work that birthed the field of *information theory*, Shannon introduced the concept of the "bit," short for binary digit, as the fundamental unit of information. A bit is a variable that can take on one of two values, often represented as 0 and 1, which we will meet again and again in this book.

In his work, Shannon connected the idea of information with uncertainty. When a message reduces uncertainty, it carries information. The more uncertainty a message can eliminate, the more information it contains. Shannon formulated this concept mathematically and developed measures for information, realizing that all types of data—numbers, letters, images, sounds—could be represented with bits, opening the door to the digital revolution. Nowadays, the concept of a bit underlies all of digital computing. Modern computers, for example, represent and process information in groups of bits.

Fast forward to the '70s, and the field of *computer science* was in full swing. Researchers devised more nuanced notions of computation, including the extension of what was "computable" to what was computable *efficiently*. Efficiency has a technical definition, which roughly means that the time to solve the problem does not "blow up" as the problem size increases. A keyword here is *exponential*. If the amount of time required to solve a problem compounds like investment interest or bacterial growth, we say it is *in*efficient. It would be computable but highly impractical to solve such a problem. The connection to physics emerged through what was eventually called the *Extended* Church-Turing Thesis, which states not that a Turing machine can't merely simulate any physical process but that it can do so *efficiently*.

Today, all computers have roughly the same architecture. A single computer can do any computational task. Some problems are hard to solve, of course. But if you can prove it is hard to solve on one computer, then you know there is no point in trying to design a new kind of computer. Why? Because the first computer can efficiently simulate the new one! This is what the Extended Church-Turing Thesis says, and it seems like common sense to the modern digital citizen. But is it true? Maybe not. When we think about simulating a physical process based on *quantum* physics, it appears that a digital computer cannot do this efficiently. Enter quantum computing.

## A brief history of quantum computing

Quantum computing was first suggested by Paul Benioff in 1980. He and others were motivated by the aforementioned interest in the physical limitations of computation. It became clear that the mathematical models of computation did not account for the laws of quantum physics. Once this was understood, the obvious next step was to consider a fully quantum mechanical model of a Turing Machine. Parallel to this, Richard Feynman lamented that it was difficult to simulate quantum physics on computers and mused that a computer built on the principles of quantum physics might fare better. At the same time, in Russia, Yuri Manin also hinted at the possibility of quantum computers, both noting their potential access to exponential spaces and the difficulty of coordinating such. However, the idea remained somewhat nebulous for a few years.

In 1985, David Deutsch proposed the *universal quantum computer*, a model of quantum computation able to simulate any other quantum system. This *Quantum* Turing Machine paralleled Turing's original theoretical model of computation based on digital information. This model of a quantum computer is essentially equivalent to the model we work with today. However, it still wasn't clear at the time whether or not there was an advantage to doing any of this. Finally, in 1992, Deutsch and Richard Jozsa gave the first quantum algorithm providing a *provable* speed-up—a quantum computer can solve in one run of the algorithm what it might take a conventional computer a growing number as the problem size gets larger. It's an admittedly contrived problem, but it is quite illustrative, and the so-called *Deutsch-Jozsa algorithm* will be discussed later.

One of the concerns at the time was how robust a quantum computer would be to noise and, indeed, any small amount of error in the algorithm destroys the computation. Of course, we could redefine the problem to be an approximate one. That is, the problem specification allows a small amount of error. The Deutsch-Jozsa algorithm then still solves the problem (it solves it perfectly, so also approximately). However, now, a digital computer can easily solve the approximate version of the problem. So, the "quantum speed-up" disappears.

Ethan Bernstein and Umesh Vazirani modified the problem in 1993 to one where small errors were allowed. They also devised a quantum algorithm that solved the problem in a single step, while any classical algorithm required many steps. In 1994, Dan Simon devised another problem along the same lines. However, this time, the quantum algorithm used to solve it provided a provable exponential speed-up. That is, as the input size of the problem grows, the best classical algorithm requires a number of steps growing exponentially, while the quantum algorithm requires, at most, a linear number of steps.

Simon's algorithm was the inspiration for perhaps the most famous quantum algorithm: Shor's algorithm. In 1994, Peter Shor presented a quantum algorithm that can factor large numbers exponentially faster than the best-known classical algorithm. Since most public-key cryptography is based on the assumed hardness of factoring, Shor sparked a massive interest in quantum computing. The race was on.

Detractors quickly dug in their heels. The most common criticism was the incredible fragility of quantum information. It was (and still occasionally is) argued that a quantum computer would have to be so well isolated from its environment as to make it practically infeasible. Classically, error correction employs redundancy—do the same thing many times, and if an error happens on one of them, the majority still tells the correct answer. However, as we will see later, quantum data cannot be copied, so it would seem error correction is not possible.

Peter Shor showed how to encode quantum data into a larger system such that if an error happened to a small part of it, the quantum data could still be recovered by suitable decoding. There was a three-qubit code and a five-qubit code and, not long after, entire families of codes to protect quantum data from errors. While promising, these were still toy examples that worked in very specific instances. It wasn't clear whether *any* required computation could be protected.

Recall that one crucial difference between bits and qubits is that the latter forms a *continuous* set. For example, 1 is a different state of information than 1.000000001, and so on. Would we need a continuous set of instructions for quantum computers for every possible transformation of quantum data? Keep in mind that a *single* instruction suffices to do any computation with bits. It's typically called the NAND ("not and") gate, and it produces an output bit of 0 when its two input bits are 1 and outputs 1 otherwise. It's amazing to think everything your computer can do, which is basically *everything*, can be done by repeatedly applying one single instruction to long lists of zeros and ones. Naively, this would be impossible for qubits. Robert Solovay and Alexei Kitaev independently proved that it isn't.

The so-called *Solovay-Kitaev theorem* showed that any quantum algorithm could be achieved with a small fixed set of instructions. This was not only important theoretically but also in practice. Quantum engineers now need only concern themselves with making a handful of operations work well—the rest can be built up from them. But while this sounds promising, there was still a loophole for the naysayers. Error correction is not perfect, even for digital electronics. (Blue Screen Of Death, anyone?) Although quantum error correction demonstrated that the most common errors occurring during the execution of these instructions can be corrected, eventually, rare errors will happen, and those will spoil the computation. Imagine you have a leak in the roof and a bucket that can catch 99% of the water. Sounds great, but eventually, the 1% that's missed by the bucket will flood the house. What you need is a bucket *and* a mop. With quantum computer errors, we had the bucket but not the mop.

The theoretical pinnacle of quantum computing research is known as the *Fault-Tolerant Threshold Theorem*. In the late '90s, several researchers independently discovered that quantum error correction allows quantum computation to happen indefinitely, so long as the rate at which errors happen is below some threshold. The exact value depends on the details of the computing model, but for the sake of argument, let's say it is 1%. What does this number mean? If you can engineer the fundamental components well enough so that errors happen less than 1% of the time, then a properly constructed quantum code will produce errors in your computation *less* than 1% of the time. In other words, you can correct errors faster than they are made, provided those errors occur at a rate below the threshold.

And that was it. Before the turn of the century, all the pieces were in place, and we just needed to wait until someone built the damn thing! But here we are, decades later, *without* a quantum computer—what gives? Is quantum technology a pipe dream? Is it too dangerous to build? Does it require access to other dimensions? And, just how exactly is the thing supposed to work? It's time we answered the real questions.

# Myth 1: Nobody Understands This Quantum Stuff

> *"Nobody understands quantum physics."*
>
> — Richard Feynman

What a quote! But, to be fair, there are so many to choose from!

*"Those who are not shocked when they first come across quantum mechanics cannot possibly have understood it."*—Niels Bohr

*"Quantum mechanics makes absolutely no sense."*—Roger Penrose

*"If it is correct, it signifies the end of physics as a science."*—Albert Einstein

*"I do not like it, and I am sorry I ever had anything to do with it."*—Erwin Schrödinger

Cute. But no one would guide their philosophical attitudes and technology predictions based on what a bunch of dead guys didn't understand, right? I mean, who would do that? Everyone, that's who!

Here's how the argument might go. While the principles we do understand have been enough to *begin* building quantum computers, it's possible that as we push the boundaries of our understanding, we might discover new principles or limitations that make quantum computing impossible or impractical. But this is just plain wrong. *We understand quantum physics exceptionally well*—so much so that we have built our entire modern society through the exploitation of our understanding of it.

## You are already using quantum technology

Don't think that all this hype about quantum technology is merely *going to* lead to a societal revolution—because it already has! Every piece of modern technology has the fingerprints of quantum physics on it. Quantum computers seem less miraculous, and their inevitability becomes more acceptable when this is understood. So, let's understand it! But first, there is something important to keep in mind.

While our precision engineering and control at the microscopic scale was a continuous process of improvement, there is still a clear distinction between *first-generation quantum technologies*, which you will learn about in this chapter, and *second-generation quantum technologies,* which include quantum computers. Once we understood the fine structure of light and matter, many new paths in understanding and engineering opened up. Yet, these did not require the manipulation of individual atoms or photons to discover, nor did they need access to the fundamental constituents to exploit. Now, we can control the world down to individual atoms. It's still quantum physics, but it provides us with more possibilities—most of which we probably don't even know about yet!

Consider the following analogy. When you bake a cake, you start with a collection of individual ingredients: flour, sugar, eggs, etc. When we gained the ability to refine and perfect the quality of these ingredients, our culinary achievements evolved from avoiding starvation to celebrity bake-offs. But, we can still only mix these ingredients together to form a general batter. This is similar to how quantum physics informed us of the constituents of matter, but we were limited to exploiting those in bulk quantities. Now, imagine being able to alter each grain of flour or sugar crystal, much like how we can now manipulate individual atoms. This would allow Gordon Ramsay to require customization of reality TV cakes to an unprecedented degree, with demands on texture, taste, and appearance beyond its recognition as food. Controlling individual atoms could allow us to design the world from scratch, producing things beyond our imagination and unrecognizable to an experience trained in a classical world.

Although that is beginning to sound like hyped-up science fiction, our ability to manipulate the world at the most fundamental scale is better seen as the natural evolution of technological progress. But what will it evolve from? Let's look at some examples.

## Transistors

Arguably, the most important technological consequence of quantum physics is the mighty transistor. If you are sitting on your mobile phone right now, you are currently sitting on a billion of these now-tiny devices. But they weren't always tiny, and the story of this technology is at least as old as quantum theory.

In the 19th and early 20th centuries, physicists discovered that silicon and some other materials had electrical properties that fit between conductors and insulators—they could conduct electricity under certain conditions but not others. Classical physics could not explain this behavior, nor could it aid in exploiting these properties. Quantum physics provided the needed theoretical framework to understand these materials. The discrete energy levels of Bohr's atomic theory generalize to the so-called *band theory of solids*. Instead of the specific energy levels for electrons in a single atom, bulk materials have "bands" and "gaps" in how electron energy can be organized in materials. Band theory suggests that conductors have many energy levels for electrons to move into, which allows electric current to flow easily. Insulators, on the other hand, have large "band gaps," preventing electron movement. Semiconductors have a small band gap that electrons can cross under certain conditions.

The detailed band structure of any material can be deduced by solving Schrödinger's equation. This is an impossible task for all but very contrived scenarios. Thus, most of the art and science of the quantum physics of solid material is in crafting useful approximations. Indeed, the sole purpose of the largest sub-field of physics—called *condensed matter physics*—is approximating quantum physics for large systems of atoms. Basically, any person or company that has sourced a product, component, or ingredient with specific properties has at least indirectly benefitted from quantum physics research. A modern transistor, for example, demands a long list of properties to function correctly.

The transistor was invented in 1947 by John Bardeen, Walter Brattain, and William Shockley at Bell Labs. While "big" compared to today's transistors, it was still only the size of a coin. Its purpose was to amplify electrical signals to replace the bulky and much less efficient vacuum tube. By using two closely spaced metal contacts on the surface of a small chunk of semiconductor material, the transistor could regulate electric current by exploiting the band structure properties of the material. After the initial demonstration, progress was rapid. In addition to replacing vacuum tubes as a necessary component in electric circuits, transistors also replaced vacuum tubes as computer switches. By 1954, the first "all-transistor" computer was built, and the rest, as they say, is history.

## Lasers

A laser used to be a LASER (Light Amplification by Stimulated Emission of Radiation), but it is now so ubiquitous that it's both a noun *and* a verb. While you don't want to be arbitrarily *lasered*, whenever you scan a barcode at the supermarket, for example, you are harnessing the power of lasers.

At the heart of laser operation is the phenomenon called "stimulated emission." This process was first described by Einstein in 1917, using the early ideas of quantum physics. Recall that an electron in an atom sits on discrete energy levels. When the electron changes levels, it absorbs or emits a photon. The energy of the photon is exactly the difference in the energy levels. *Spontaneous* emission happens when atoms in high-energy states randomly drop to low-energy states. All the light you see, and all the light ever seen before the 20th century, was due to atoms randomly changing energy levels. Einstein suggested that an atom in an excited state (with its electron at a high energy level) could be *stimulated* to drop to a lower energy state with a photon matching the energy level difference, thereby creating two identical photons.

Despite Einstein's theoretical work, it wasn't until the 1950s that scientists were able to build devices that took advantage of stimulated emission. The first was developed by Charles Townes and his colleagues at Columbia University in 1954, though it still wasn't practical and only amplified microwave frequencies rather than visible light. A proper laser was first demonstrated in 1960 by Theodore H. Maiman at Hughes Research Laboratories. It was red, by the way. But while impressive, it was still famously described as a "solution looking for a problem." Interestingly, most applications of lasers today solve problems no one even dreamed of having in 1960.

Just like the transistor, the laser has found many uses. Today, lasers are used in a wide variety of fields, from medicine and telecommunications to manufacturing and entertainment. They are used to cut steel, perform delicate eye surgeries, carry information over fiber-optic cables, read barcodes, read and write music and video onto plastic discs, and create mesmerizing light shows. Our modern world could not exist without the laser, which will continue to pay dividends when applied to second-generation quantum technology.

## Atomic clocks

If you've ever used GPS (Global Positioning System) to navigate, which is nearly impossible not to unless you carry no devices and drive yourself around in a twenty-year-old vehicle, then you've indirectly used an *atomic* clock. Since it has the word "atom" right in it, you know it has something to do with quantum physics.

The development of atomic clocks was made possible by understanding the detailed internal quantum nature of atoms. Electrons in atoms occupy discrete energy levels and can transition between these levels by absorbing or emitting specific frequencies of light. That's old news to us now. What's new is that these levels can be manipulated with electric and magnetic fields—otherwise, single energy levels can be *split*. This splitting can even come from electric and magnetic fields generated from *within* the atom—electrons carry an electric charge, after all! The natural splittings are subtle but reveal what was called the *fine* and *hyperfine* structure of atoms. In the 1930s, Isidor Rabi developed the technique of magnetic resonance, enabling the precise measurement of these features.

Since the energy level structure is a fixed property of atoms, and the frequency they omit would create a very precise "ticking," Rabi later suggested that atoms could be used to define an extremely stable clock. In 1949, the United States National Bureau of Standards built the first atomic clock using ammonia. But it was the cesium-based clock, developed in 1955 at the National Physical Laboratory in England, that truly revolutionized timekeeping.

Cesium-based atomic clocks work on the principle of measuring the frequency of cesium atoms when they transition between two specific energy levels. The lowest energy hyperfine transition of cesium atoms oscillates exactly 9,192,631,770 times per second, and this frequency was so constant and reliable that in 1967, it was adopted as the new standard for the second, replacing the previous standard based on the Earth's orbit. Our definition of time itself is based on quantum physics.

These incredibly precise timekeepers have been used in a variety of applications, from synchronizing telecommunications networks like the internet to testing the predictions of Einstein's theory of relativity. However, perhaps their most well-known application is in the Global Positioning System. Each of the 24 satellites in the GPS constellation carries multiple atomic clocks on board. The tiny differences in the time signals sent by these clocks, caused by their different distances from the receiver, allow the receiver's position to be triangulated within a few meters. Measuring distances is hard, but since light has a constant speed, distance can be inferred by how long it takes to travel—provided you can accurately measure time.

The atomic clock is a prime example of a quantum technology that has become indispensable in our modern world. Whether it's enabling global navigation or the internet, the atomic clock demonstrates that we are living in the quantum technology revolution—the first one, anyway.

## Magnetic resonance imaging (MRI)

Magnetic Resonance Imaging, commonly known as MRI, is a powerful tool in the medical world, providing detailed and non-invasive images of soft tissues in the body—something traditional X-rays struggle with. The technology behind MRI is rooted in quantum physics, drawing directly from the principles of magnetic resonance mentioned with atomic clocks.

One of the hallmarks of quantum physics was the discovery of *spin*, a property internal to subatomic particles that forces them to act like tiny magnets. Individually, they are very weak and, when surrounded by others aligned in random directions, are impossible to detect. However, they will align themselves with a strong enough magnet, and that's where the giant superconducting coil magnets of an MRI machine come in.

In MRI, the focus is mainly on the spin of hydrogen due to the abundance of water in our bodies. When placed inside the MRI machine's strong magnet, these spins align with it. A signal is then applied perpendicular to this magnetic field, causing the spins to tip away from their aligned state. When the signal is removed, the spins return to their original alignment, but in the process, they emit signals back. These signals, from the collective relaxation of billions upon billions of tiny magnets deep within hydrogen atoms, are what the MRI machine captures and converts into images.

The densities and types of tissues result in varied relaxation times, meaning the signals received from various tissues will differ. This difference allows for a contrast in the images and enables the detailed visualization of organs, tumors, and other structures.

Without understanding the spin properties of atomic nuclei, the development of MRI would not have been possible. Today, MRI is used globally to diagnose a myriad of conditions, from brain tumors to joint injuries, showcasing yet another practical application of quantum physics in our everyday lives.

## And many more…

While we have covered some of the most significant applications of quantum physics in technology, there are, of course, many more. For instance, nuclear energy and, unfortunately, weapons are widespread applications relying heavily on the principles of quantum physics via nuclear physics. Without knowledge of the internal workings of atoms, we would not have the over 400 nuclear reactors providing 10% of the world's electricity, nor would we have the over 200 research-grade reactors that also produce radioactive material for industrial and medical purposes.

Speaking of which, beyond MRI, quantum physics has had a profound impact on medical technology. Positron Emission Tomography (PET), for example, uses the obscure-sounding *anti-particle* to the electron to construct images of the body's internal processes. A PET scan introduces a small amount of radioactive material into the body, which emits radiation as positrons. When they meet electrons, the emitted positrons annihilate and produce two gamma rays in opposite directions. The PET scanner detects these gamma rays, infers the source, and hence maps out the journey of the radioactive material within the body.

Semiconductor technology has led to numerous advances over the years beyond the ubiquitous transistor. While reliant on quantum physics for many aforementioned reasons, quantum knowledge is doing double duty in *tunnel diodes*. In these devices, electrons can "tunnel" through energy barriers instead of requiring to go over them, as one might expect from classical physics. Various versions of these are used to improve everything from Christmas lights and solar cells to lasers and thermometers.

Moreover, quantum concepts and mathematics have begun to find applications in fields far removed from traditional physics. Quantum biology, for instance, applies quantum principles to biological processes, investigating phenomena like photosynthesis and bird navigation, where a classical description alone may be insufficient. "Quantum" finance borrows mathematical techniques from quantum mechanics to model financial markets and to understand their seemingly random fluctuations better.

Quantum physics is far more embedded in our everyday lives than we might initially realize. From the electronic devices we use to navigate our world to the medical technologies that help diagnose and treat illnesses to the mathematical models that power our financial systems—the principles of quantum mechanics are an integral part of our intentionally engineered modern world. And this doesn't even touch on the second-generation quantum technologies that are on the horizon. As our control of quantum systems continues to improve, more applications will come, so this all begets the question, *what exactly don't we understand about quantum physics?*

## Quantum teleology

Quantum physics is almost always taught chronologically. Indeed, I just did that in the previous chapter. You read about a long list of 20th-century scientific heroes who uncovered the wild and untamed world behind our fingertips. The story had modest roots in Planck's 1900 hypothesis that energy is discrete. Though we didn't need to make it that far for the purpose of introducing quantum computers, the standard tale of quantum physics usually crescendos with John Bell's work on "spooky" entanglement in the 1960s. Today, as the story goes, we are on the cusp of the yet-to-be-written second quantum revolution.

Along the way, the standard story tells of a piece of great machinery that was simultaneously being created next to quantum *physics*, called quantum *mechanics*, which allowed graduate students to blindly turn the mathematical crank to make predictions about newer and more extreme experiments. It is often said that generations of physicists would "shut up and calculate" to earn their degrees and professorships to eventually repeat the program again with the next cohort. The wild horse of quantum physics had apparently been stabled but not tamed.

When Richard Feynman made the casual remark about "understanding," he was not only opening his now-famous lecture on quantum physics, he was inadvertently venturing into the realm of philosophy, which he famously derided. In epistemology—the study of knowledge itself—understanding is not just a mechanical mastery but a rich concept that probes the underlying meaning and connection between ideas. This brand of philosophy seeks to clarify the very nature of knowing and being. As you might have expected from such a grandiose task, there's more disagreement than agreement amongst philosophers on even the definition of the word.

On the other hand, we all have some intuitive notion of understanding. In the everyday world, understanding might be likened to knowing how to ride a bike, but in physics, it's about grasping the forces that make the bike move. You might not know the physics behind balance and motion, but you come to "understand" how to ride it through feel and experience.

When we talk about physicists' understanding, we're stepping into a workshop where the universe's machinery is laid bare. A physicist strives to see the gears and levers behind phenomena, aiming for an intuitive grasp of why things happen as they do. This is not merely knowing the equations but being able to feel them, like instinctively leaning into a turn.

However, in the realm of quantum physics, the rules of the game seem to change. Understanding here is like trying to catch smoke with your bare hands—it slips and dances between your fingers. The intuitive mechanics known in classical physics fade, and in their place are abstract mathematical objects and formulas. You can follow them, learn them, even use them, but a concrete mechanical understanding of them appears impossible. Imagine riding a bike—built by no one—with all its mechanisms hidden and impossible to reveal.

If all the mixed messages about quantum physics confuse you, I want you to erase everything you know about it and memorize the next paragraph.

Quantum physics is a branch of science that describes highly isolated systems—things that don't interact randomly with other stuff around them. Traditionally, these are small, like atoms, but now we can engineer artificial systems under high isolation. Anything that is extremely isolated requires quantum physics to be described accurately. The information such things encode is quantum information. If you attempt to use classical physics or classical information to make predictions or statements about such things, you may end up being very wrong.

Quantum physics does not tell us what reality lies beyond our experience. It only tells us that, whatever it is, *reality* is nothing like the mechanical worldview we have come to take for granted in realms where it works. Thus, if "understanding" demands such an explanation of the world—displaying the causes and effects that make it go—indeed, no one yet has it.

But this myth is not about philosophy because this is a book about technology. So, we will stick to the everyday use of words. That is, using a tool is exactly how you come to understand how it works. Generally, people stop demanding explanations of things they are familiar with. We will eventually become so familiar with quantum technology that understanding will be a word reserved for one's ability to successfully navigate its use. In much the same way that my inability to "understand" the point of TikTok does not stand in the way of it being a successful app, our lack of "understanding" quantum physics does not stand in the way of building quantum computers.

If Feynman were alive today, I think he'd contextualize this quote better. He might say, "Quantum mechanics cannot be understood using classical physics and information—it must be understood in the language of quantum information." Or, perhaps, he'd be pithier and say, "Nobody *speaks* quantum information." Indeed, nearing the end of his career, he said, "Nature isn't classical, dammit, and if you want to make a simulation of nature, you'd better make it quantum mechanical, and by golly, it's a wonderful problem, because it doesn't look so easy." The workshop he said this at in 1981 is often considered the birthplace of quantum information and computation.

# Myth 2: Qubits Are Zero And One At The Same Time

> *"Well, some go this way, and some go that way. But as for me, myself, personally, I prefer the shortcut."*
>
> — Lewis Carroll

You probably think that a qubit can be 0 and a 1 at the same time. Or that quantum computing takes advantage of the strange ability of subatomic particles to exist in more than one state at any time. I can hardly fault you for that. After all, we expect places like the New York Times, Nature, Science, New Scientist, Time Magazine, and Scientific American, among many others, to be fairly reputable sources, right? Apparently not. Nearly every popular account of quantum computing has this "0 and 1 at the same time" metaphor.

I say metaphor because it is certainly not literally true that the things involved in quantum computing are 0 and 1 at the same time. Remember that computers don't actually hold 0s and 1s in their memory. Those labels are just for our convenience. Each bit of a digital computer is a physical thing that exists in two easily distinguishable states. Since we use them to perform logic, we could also have labeled them "true" and "false." Now it should be obvious—"true and false at the same time" is just nonsense. (In formal logic, it is technically a false statement.)

The problem is even worse, though, because not only is "0 and 1 at the same time" a gross oversimplification, but it is also a very misleading analogy.

## If 1 = 0, I'm the Pope

Me and the Pope are clearly two people. But assume 1 = 0. Then 1 + 1 = 0 + 1. Since 2 = 1, me and the Pope are also one person. Therefore, I am the Pope.

You see where the problem is, right? From a false premise, any conclusion can be proven true. This little example was humorously pointed out by the famous logician Bertrand Russell, though he wasn't talking about qubits. However, we can clearly see that starting with a blatantly false statement is going to get us nowhere.

Consider the following logic. First, if a qubit can be 0 and 1 at the same time, then two qubits can be 00 and 01 and 10 and 11 at the same time. Three qubits can be 000 and 001 and 010 and 011 and 100 and 101 and 110 and 111 at the same time. And… well, you get the picture. Like mold on that organic bread you bought, exponential growth!

The number of possible ways to arrange $n$ bits is 2 to the power of $n$—$2^n$, a potentially big number. If $n$ is 300, then 2 to the power of $n$ is $2^{300}$, which is more than the number of atoms in the universe! Think about that. Flip a coin just 300 times, and the number of possible ways they could land is unfathomable. And 300 qubits could be all of them *at the same time*. If you believe that, then it is easy to believe that a quantum computer has exponential storage capacity and power. That would be magic. Alas, this is not how qubits work.

Clearly, we just need to get "0 and 1 at the same time" out of heads.

## Escalating quickly

Let's skip ahead to the end just for a brief moment. I'm going to tell you what qubits actually are, if only so that I can say I never held anything back. If you take a formal university subject in quantum computing, you will learn that qubits are *vectors in a complex linear space*. That sounds complicated, but it's just jargon. *Vectors* are lists of numbers, *spaces* are collections of vectors that are *linear* because you can add vectors together, and the word *complex* refers to numbers that use the square root of -1, which is also called an *imaginary* number.

Vector spaces are used everywhere, from finance to data science to computer graphics, and despite the term "imaginary" associated with the square root of -1, complex numbers have very real applications as well. Their beauty lies in their ability to simplify and streamline otherwise complicated mathematical problems across various disciplines, from fluid dynamics to electrical engineering. So, the tools themselves are not mysterious. The rules for how these things are used in quantum computing are not complicated either. However, this *is* where I will stop short of turning this into a math textbook because I'm sure you're here to read words and sentences rather than symbols and equations.

In short, a qubit is represented not by "0 and 1 at the same time" but by two complex numbers. These numbers take on a continuum of values, so they are indeed much more versatile than the binary option afforded to a single bit. However, they do come with limitations that prevent them from being a computational panacea. If confusion sets in at any point, remember that qubits are lists of complex numbers, and there is a very solid mathematical framework for dealing with them.

## Putting the word quantum in front of everything

In the Marvel movie *Ant-Man and The Wasp*, the term "quantum" is bandied about so liberally that Paul Rudd (playing Ant-Man) nearly breaks the fourth wall to ask on behalf of the audience, "Do you guys just put the word quantum in front of everything?"

While this was meant to be a joke in the context of the movie's dialogue, the answer in the real world is an emphatic *yes*. It's an inside joke that to be a quantum information theorist amounts to opening a classic textbook on information theory and literally placing the word "quantum" in front of every definition and theorem. Though we have quantum information, quantum entropy, quantum channels, quantum coding, quantum this, and quantum that, we didn't hold on to "quantum bits." In fact, we did away with it rather quickly.

Qubits were introduced in a 1995 physics paper titled *Quantum Coding* by Benjamin Schumacher as follows.

*"For our elementary coding system we choose the two-level spin system, which we will call a 'quantum bit' or qubit. The qubit will be our fundamental unit of quantum information, and all signals will be encoded into sequences of qubits."*

Boom. Qubits burst on the scene with authority! But wait…what's this buried in the Acknowledgments section?

*"The term 'qubit' was coined in jest during one of the author's many intriguing and valuable conversations with W.K. Wootters, and became the initial impetus for this work."*

Ha! The lesson? Always see a joke through to the end.

## Writing in qubits

Usually, you will see qubits "written" with a vertical bar, |, and a right angle bracket, ⟩, which come together to form something called a "ket." There is always something "inside" the ket, which is just a label. Just as variables in mathematics are given letter symbols ("Let $x$ be…" anyone?), an unspecified qubit is typically given the symbol $|\psi\rangle$. The notation, called *Dirac notation*, is not special among the various ways people denote vectors, but physicists have found it convenient. Since quantum computing was born out of this field, it has adopted this notation.

The other important thing to note is the use of the word *state*, which is confusingly overloaded in both physics and computation. The object $|\psi\rangle$ is often called *the* state of a qubit or that *the* qubit is *in* the state $|\psi\rangle$. Sometimes $|\psi\rangle$ is taken to be equivalent to the physical device encoding the information. So, you'll hear about *the state of physical qubits*, for example. This is more of a linguistic convenience than anything else. While there is nothing wrong with using this short-hand in principle, it is what leads to misconceptions, such as things being in two places at once, so caution is advised.

Imagine I hand you a USB stick with a message. It might be said that I've given you several bits—as if the information were a physical quantity and the USB stick *is* the bits containing my message. But, again, that's just a convenient and economical way to speak about it. Really, I have given you a physical device that can represent a bunch of binary options. I encode my message into these options, and you decode the message by looking at it. The *state* of the USB stick can be described by my message, but it is not literally my message.

The same logic applies to qubits. I can encode qubits into physical devices. Out of convenience, we say those devices *are* qubits. It may then seem like the statement "the qubit is in the state $|\psi\rangle$" implies that $|\psi\rangle$ is a physical quantity. In reality, $|\psi\rangle$, which is quantum information, can only describe the state of the physical device—whatever current configuration it might be in. That configuration might be natural, or it might have been arranged intentionally, which is the process of encoding the information $|\psi\rangle$ into the physical device.

Much like the process of encoding bits into a USB stick is referred to as "writing," encoding qubits into some quantum device is the writing of quantum data. In the old parlance of quantum physics, this is the same as "preparing" a quantum system, where $|\psi\rangle$ summarizes the repeatable laboratory procedure to bring a physical system into a particular configuration. In the past, this was for the purpose of experimentation. Now, it is done for the purpose of computation.

All this is to say that, within any specialized discipline, people are sloppy with their jargon. The trouble with quantum computing is that the sloppy jargon is the only thing that leaks out of a field that remains specialized. When these phrases are combined with our everyday conceptions of the world, we get weird myths.

## Superposition

It's been mentioned that a qubit is simply a pair of numbers. There are infinitely many pairs of numbers, but also some special ones. For example, the pair (1,0) and the pair (0,1) are pretty special. In fact, they are so special they are given their own symbols in Dirac notation: |0⟩ and |1⟩. Among the myriad of options, this choice was made to keep the connection to the bits 0 and 1 in mind.

The other pair of numbers that usually gets its own symbol is (1,1). The symbol for this pair is |+⟩, and it's often called the "plus" state. Why plus? Ah, we've finally made it to *superposition* and the origin of "0 and 1 at the same time." This is the only bit of math I'll ask you to do. What is (1,0) + (0,1)? That's right, it's (1,1). Writing this with our symbols, |+⟩ = |0⟩ + |1⟩.

The qubit is not in the |0⟩ state, nor is it in the |1⟩ state. Whatever we might do with this qubit seems to affect both |0⟩ and |1⟩ at the same time. So, it certainly does *look* like it is both 0 *and* 1. Of course, the reality is more subtle than that.

How would one physically encode the state |+⟩? Naively, the equation suggests first encoding 0, then encoding 1, and,

finally, adding them together. That seems reasonable, but it's not possible. There's never any addition happening in the physical encoding or processing of qubits. The only reason it is written this way is out of convenience for scientists who want to write the states of qubits on paper. Taking a state $|+\rangle$ and replacing it with $|0\rangle + |1\rangle$ is an intermediate step that students learn to perform to assist in calculations done by hand. A quantum computer could not do this, nor would it need to. A quantum computer holds in its memory $|+\rangle$, full stop.

Any pair of numbers that is not (1,0) or (0,1) can be written as a sum of the two of them. Such states are called *superposition states*, and the nomenclature gets distorted into phrases like "the qubit is in superposition." You can probably imagine, or have already seen, many misinterpretations of such a statement. The words tempt us to think that a qubit in superposition is in multiple states simultaneously. However, the simplest true interpretation, though not all that compelling, is just "a qubit in superposition is not one that is encoded as either $|0\rangle$ or $|1\rangle$." This is exactly why I never get invited back on quantum hype podcasts…

At this point, you may be wondering why we use the labels 0 and 1 in the first place if they are so prone to confusion.

There is a good reason for it, and it comes when we attempt to *read* qubits.

## Reading quantum information

Sorry, but *you can't read quantum information.*

In a world where classical physics reigns supreme, reading data is straightforward. If you've saved a document on your computer, when you open it later, you expect to find the same content. Moreover, reading the content amounts to directly perceiving the symbols encoding bits of information. It is so obvious and intuitive we barely give it a second thought. But when it comes to quantum data, things are much different.

In classical physics—and everyday life—measurement is the process of determining the value of pre-existing properties of things, like weight, dimensions, temperature, and so on. With a well-calibrated instrument, we can "read off" what was always there. We can encode information into these properties and, if all else remains the same, decode the information later.

Now, think about an atom for a moment. It's tiny. Unimaginably tiny. There are mountains of irrefutable evidence that atoms are real, even though no one has ever *seen* an atom. What we see with our naked eyes is information on the displays of large instruments. But that information is classical, represented in the digital electronics of the device as bits. In short, any attempt to gain information from a quantum system results in bits, not qubits. We cannot simply "read off" the state of a qubit.

In physics, there is plenty more jargon surrounding this, including *measurement*, *observables*, and *collapse*—none of which is important for quantum computing. All we need to understand is that reading quantum data results in classical data—*n* qubits of information produce *n* bits of information when read.

As an example, take some qubit of information—some $|\psi\rangle$ from earlier. Suppose it is encoded into the energy levels of an atom. Any attempt to "read" the atom by, say, measuring the amount of energy it has will result in a binary outcome. (The atom will decay and give off a photon of light or not.) That's one bit of information. Since $|\psi\rangle$ is specified by two continuously varying numbers, one bit is not nearly enough to resolve which pair it is. In other words, you can't read quantum information.

## Quantum measurement

In the previous example, an atom was imagined to encode a qubit of information in its energy state. When read, one of two outcomes will occur. If the atom is in the high energy state, it will release that energy as a photon. But, now it has no energy, so it must be in the low energy state. While there are many clever ways to write and read qubits from physical systems, none of them can avoid this situation. Some of the verbs that have been associated with the outcome of a read qubit are destroyed, deleted, collapsed, and other gruesome-sounding terms. A more straightforward way to say it is simply that the physical system no longer encodes the quantum data.

Going from the classical notion of measurement to the quantum one is a huge physical and philosophical leap and something scientists and philosophers still debate about. So, I'm not going to attempt to give a complete answer to *why* reading qubits works this way, but I'll give you the gist of it. To "measure" even large classical systems is often invasive. Things like biopsies make that obvious. A less complicated example is tire pressure. Given a tire, we assume the air inside it has some pressure—some fixed value for that property of air, which is true of the air whether we attempt to measure it or not. However, actually measuring the air pressure requires opening the value to move a needle on some gauge. By letting at least a little bit of air out, we've changed the value of the very thing we were attempting to measure.

You might intuit that the more you learn about a system, the more you change it. Quantum physics is what you get when you take that idea to the extreme. We can manipulate quantum objects without disturbing them, but then we would gain no information from them. We can eek a small amount of information at the cost of little disturbance, but extracting the most information possible necessitates maximum disruption. Imagine a tire pressure gauge that lets out *all* the air and only reports whether the tire was previously full or not. You learn a single bit of information and are always left with a flat tire. Though that's a good analogy, I promise that quantum computers are more useful than it sounds.

## A game of chance

So far, we have that qubits encode quantum data but can only reveal a smaller amount of classical data. But there's something that should be nagging at you—surely the outcome has to depend somehow on the quantum information $|\psi\rangle$. Otherwise, what's the point? Indeed. However, it's not the *outcome* itself that depends on $|\psi\rangle$, but the *probability*.

Quantum physics is not a deterministic theory. It gives us very accurate predictions of probabilities for the possible outcomes of experiments, but it does not tell us which outcome will happen on each occasion. That is, when we read a qubit, the classical bit we receive is *random*. Recall the plus state from before, $|+\rangle = |0\rangle + |1\rangle$. When read, it will produce the bit 0 or the bit 1 with equal probability. You might call it a *quantum* coin—a perfectly unbiased random event. Indeed, this is the basis of commercially available QRN, or Quantum Random Number, generators.

It's going to be our mantra by the end of this book, but a qubit is a pair of numbers. Let's give the pair symbols (*x*,*y*). If either of the pair is zero, the result of reading the qubit will be deterministically 0 or 1. That is, reading the state $|0\rangle$ results in 0, and reading the state $|1\rangle$ results in 1. All other states have some unavoidable randomness. If *x* is larger than *y*, the outcome is biased toward 0. There's a mathematically precise rule for this called the *Born Rule* in quantum mechanics, but it requires too much symbolic baggage to present here. Besides, you've got the general idea.

## Why care about coin tosses?

If quantum computing is based on such uncertainties, how can it be useful? This is where the richness of quantum *algorithms* comes into play. Quantum algorithms are designed around the uncertainties associated with reading qubits. The task of an algorithm designer is to amplify the probabilities associated with correct solutions and minimize the probabilities of incorrect ones. So, even if individual qubit measurements are uncertain, quantum algorithms as a whole guide the computation toward a useful outcome. This is the topic of the next myth, so there's more to come on algorithms.

To summarize, a qubit is not "0 and 1 at the same time" but rather is described by a pair of numbers representing its state. This mathematical framework is rich and allows for a wide array of manipulations beyond what can be done with classical bits. Writing and reading qubits involve encoding them into and decoding them from physical systems, but with significant differences compared to classical bits. Most notably, "reading" a qubit is a random event that yields classical information and destroys the quantum information. The qubit state influences the outcome's probability but isn't fully revealed in the process. The challenge, fascination, and potential power of quantum computing lie in navigating these intricacies to perform useful computations.

# Myth 3: Quantum Computers Try All Solutions At Once

> *"There are two things you should remember when dealing with parallel universes. One, they're not really parallel, and two, they're not really universes."*
>
> — Douglas Adams

The most fantastical, and hence the most popular, explanation of quantum computing is that it happens simultaneously in parallel worlds. While this is based on the speculations of Very Serious Scientists, it's not realistic and leads to misconceptions such as the idea that a quantum computer tries all the solutions to a problem at the same time so it can instantly produce the answer.

The idea of parallel universes has long captivated human imagination. From science fiction novels to blockbuster movies, the thought of coexisting realities where alternative versions of ourselves live out different destinies is undeniably intriguing. So, when quantum computing—a field riddled with complexities and counterintuitive principles—emerged, it's unsurprising that the allure of the parallel universe concept became a go-to explanation for many.

Within this captivating narrative, quantum computers were hailed as miraculous machines that could tap into these alternate realities. It was suggested that, perhaps, these computers operated simultaneously across numerous universes, hence their unparalleled speed and power. Such an idea is not only a testament to our propensity for wonder but also an indicator of how complex quantum physics is to the uninitiated.

## Many worlds, one origin

The "quantum tries all solutions at once" myth derives from the *Many-Worlds Interpretation* (MWI) of quantum physics, which can be traced back to Hugh Everett III's 1957 Ph.D. thesis, *On the Foundations of Quantum Mechanics*. However, it was mostly ignored until 1970, when Bryce DeWitt resurrected it in an article, *Quantum Mechanics and Reality*, appearing in Physics Today. Since then, a growing number of physicists have subscribed to the idea, many referring to themselves as *Everettians*.

Most popularizations of the MWI focus on the metaphors of a universe that "branches" into "parallel" worlds. This leads to all sorts of confusion. Not only can you waste your money on a Universe Splitter iPhone app (which definitely doesn't split anything), but physicists even argue amongst themselves at the level of these metaphors. Let's call this kind of stuff Metaphorical Many-Worlds and not discuss it further. Is there a better way to think about the Many-Worlds Interpretation than this? Yes—and the first thing we are going to do is stop calling it that. Everett's core idea was the *universal wave function*. So what is that?

I briefly introduced the wave function as the symbol $|\psi\rangle$ in the previous myth. Quick recap: a wave function is a mathematical variable (like "*x*" from algebra class, but dealing with complex numbers) used to calculate what will be observed in experiments. In the context of quantum computers, it's the quantum information (the qubits) encoded into some physical system. It's all the information needed to predict what will happen, summarized in the most succinct way possible.

The Schrödinger equation dictates how the qubits, or the wave function, change in time, except, of course, when we attempt to read that information—recalling that reading qubits destroys the quantum data. In other words, there are two rules in quantum physics for how qubits change, and when to apply them is arbitrary and at the discretion of the user of the theory. This bothers all physicists to some extent but bothered Everett the most.

## And one wave function to rule them all

The executive summary of Everett is this: quantum theory is consistent without any rule about reading qubits if we consider the quantum information that describes the entire universe—the universal wave function. This wave function evolves according to Schrödinger's equation always and forever.

This is where all Everettians start. Popular science writer and fervent MWI supporter Sean Carroll calls it "Austere Quantum Mechanics" for its apparent beauty and simplicity. One state, one equation—all's right with the world. There's also just one problem—it doesn't fit at all with our experience of reality. We don't experience the world as quantum things—being in superposition and whatnot—we experience a definite classical world. We really do experience the effects of reading quantum data. So where does my experience of a qubit in superposition only revealing a single bit fit into the universal wave function?

Let's go back to superposition. Remember that $|0\rangle$ is a qubit of information in a definite state—reading it deterministically produces the bit 0. We might as well call it a "classical" state. The same goes for |1⟩. On the other hand, the state |0⟩ + |1⟩ is *not* a classical state as it cannot be encoded into something that only holds bits. If we expand on these descriptions to include more possibilities, the amount of information grows. Luckily, our notation remains succinct. Let's say that |world 0⟩ is a definite classical state of the entire universe, as is |world 1⟩, and they differ *only* by one bit (the outcome of reading a single qubit).

In classical physics, "adding worlds" has no meaning, but in quantum physics, |world 0⟩ + |world 1⟩ is a perfectly valid state. If we imagine creating a definite classical state for each of the mutually exclusive events possible and then adding them all up, we will end up with one big superposition state—the universal wave function. Now, we don't need to write that all down to interpret it—the simple two-state model suffices, so we'll stick with that. And, it seems to say not that there is a single world that suddenly and randomly jumps upon reading a qubit to either |world 0⟩ or |world 1⟩, but two worlds that exist simultaneously.

## Quantum Dad

David Deutsch is often referred to as the "father of quantum computing." As noted in the brief history presented in the introductory chapter, Deutsch conceived of a model of computation called a universal quantum computer in 1985.

Deutsch's motivation was to find "proof" that MWI is correct. Deutsch is clearly a proponent of the MWI, and he has speculated exactly that which we are referring to as a myth. In his view, when a quantum computer is in a superposition of states, each component state corresponds to a "world" in which that particular computational path is being explored. He dubbed this "quantum parallelism" and suggested that the physical realization of a quantum computer would be definitive experimental evidence of MWI.

Here's the basic idea in language we have already introduced: if something acts on the superposition state |world 0⟩ + |world 1⟩ as a whole, it seems like it simultaneously acts in both worlds. In his seminal paper, he detailed a small example (now called Deustch's algorithm) that makes this logic more concrete.

The first thing to note is that algorithms of any type are recipes that solve all instances of a more generic problem. Recall long division—it was not a sequence of steps that worked for only one problem, but *any* division problem. To describe Deutsch's algorithm then requires that we understand the problem it is meant to solve.

Consider a simple one-bit computer that accepts a single bit as input and produces a single bit as output. There are not many programs we can run on such a computer. The program could do nothing and return the same bit it received. Or, it could switch the value of the bit (0 becomes 1 and 1 becomes 0). It might also ignore the input bit entirely. It could produce 0 no matter what the input was, or it could produce 1 no matter what the input was. And that's it. Those four options are the only possible ways to manipulate a single bit. We can split these four programs into two categories: the pair whose outputs added together are odd, and the pair whose outputs added together are even. Given a program, Deutsch's algorithm tells you which category it belongs to.

If you were given a digital computer with an unknown program, you would expect that you would need to use it *twice*—once for each possible input bit—and add the outputs together. However, Deutsch showed that if you could input a qubit into the computer, and that qubit was in a superposition state, you only need to run the program once. He later showed, with Richard Jozsa in 1992, that the same is true no matter how many input bits there are. In other words, this is a problem that a digital computer requires exponentially many uses to solve, but only a single use of a quantum computer. It seems the quantum computer has run the program on all inputs simultaneously.

Deutsch then asked if all of that computation is not done in parallel universes, where could it possibly happen?

## It happens here

The key to Deutsch's claim is a mismatch in resources. It doesn't take that big of a problem before all *possible* solutions outstrip the total number of things in the entire (single) universe we could use to encode bits. Therefore, the quantum computer must be using resources in other universes.

The problem with this logic is that it discounts quantum information altogether. Sure, it takes exponentially many bits to encode the same number of qubits, but we also have the qubits here, in this single world. Of all people, those who envision a cosmos of countless classical universes seem to lack the imagination to picture a single universe made of quantum stuff instead of classical stuff.

Any argument that quantum computers access parallel worlds or more than three spatial dimensions relies on circular logic that presupposes the objective reality of each component of a superposition state. In other words, they use the MWI to prove the MWI.

## Naive quantum parallelism

Now, even if you still want to believe the MWI to be the one true interpretation of quantum physics, its implications for quantum computing are just not useful. In fact, they appear to be dangerous. Computer scientist Scott Aaronson, probably the most famous popularizer (and tied for the most curmudgeonly), bangs on this drum (out of necessity) in every forum he's invited to, *and* he appears to be as sympathetic to MWI as you can get without officially endorsing it.

The most obvious logical step from quantum parallelism is that quantum computers try every solution to a problem simultaneously. This runs into two major problems, as Aaronson points out. First, it implies that quantum computers can efficiently solve some of the hardest problems we know of (a famous example being the *Traveling Salesman Problem*). However, Grover's algorithm (which we will see later) is the only known quantum algorithm that can be applied generically to such problems, and it provably has only a modest advantage. Technically, unless some new quantum algorithm appears that would upend our understanding of computer science, physics, and philosophy, quantum computers will not be able to solve such problems efficiently.

The second issue with naive quantum parallelism is that supposing a quantum computer *could* access alternative universes, it seems to do so in the most useless way possible. Rather than performing exponentially many computations in parallel and combining the results, it simply returns a random answer from one of the universes. Of course, actual quantum algorithms don't work that way either. Instead, algorithms manipulate the coefficients of superpositions (whether there is a "plus" or "minus" between $|0\rangle$ and $|1\rangle$) so that "correct" answers are returned when the quantum data is read. Crucially, quantum computers can only do this for very specific problems, suggesting that the power of quantum data is not access to parallel worlds but simply a matching of a problem's structure to the mathematics of quantum physics.

## Einstein to the rescue

Contrary to popular belief, Einsteinian relativity does not render Newtonian gravity obsolete. In fact, we probably use and refer to Newton's idea *more* now than we did before general relativity came along, though that is mostly just because there are more scientists and engineers trying to launch things around than there were a century ago. Occasionally, we even appeal to Newtonian gravity to explain or interpret general relativity. Consider the common technique of placing a bowling ball on stretched fabric to simulate the warping of spacetime. As the ball pulls the fabric down, we can appeal to our intuition from Newtonian gravity to predict what will happen to smaller balls placed on the fabric. This is analogous to when we try to interpret quantum computers through the lens of classical computers—we are explaining the new idea in the context of the older ones. However, we can also explain older ideas through the lens of newer ones, which often have much more explanatory power.

Suppose you've just had a riveting lecture on general relativity—which ought to have blown your mind and upended your conception of reality—and are now wondering how the old ideas of Newton fit into the new picture. Einstein asked us to imagine being in a rocket ship in the dead of space, with no planets or stars nearby. Of course, you would be floating inside your rocket ship, feeling weightless. Now, imagine the rocket ship started accelerating forward at a constant rate. The ship would move forward, but you would remain still until the floor of the ship reached you. At that point, the floor would provide a constant force pushing on you. You could "stand up" and walk around on the floor, which now gives you the sensation of weight. In fact, you have no way of knowing whether the rocket ship is accelerating in empty space or is simply standing up, completely still, on Earth. The "feeling" of gravity is just that, a feeling. In other words, gravity is a "fictitious" force like the "centrifugal force" keeping water in a bucket that swung around fast enough. Once you take examples like this on board, you tend to understand *both* Einsteinian and Newtonian gravity better. Can we do the same for quantum computers?

A quick recap. A qubit is two complex numbers—one associated with 0 and the other with 1. Two qubits are represented as four complex numbers associated with 00, 01, 10, and 11, and so on it goes. A large number of qubits is an exponentially large list of complex numbers, each one associated with a possible ordering of bit values. Ten qubits have $2^{10}$, or 1024, complex numbers, one of which is associated with 0000000000 and another with 1001011001, and so on. A quantum computation manipulates these numbers by multiplying and adding them with other numbers. One classical interpretation is that the quantum computer is doing classical computation on each of these bit values simultaneously. That's simple and neat but, as noted above, quickly leads to misconceptions. Let's consider then taking this quantum computation description for granted and ask how classical computers fit in.

An interesting generalization of digital computers includes randomness, which you can imagine comes about by flipping coins to decide the input of the program. How these "probabilistic bits" are described is actually remarkably similar to qubits. One probabilistic bit is two numbers—the probability of 0 and the probability of 1. Two probabilistic bits are four numbers representing the probability of 00, 01, 10, and 11. Just like with qubits, ten probabilistic bits have $2^{10}$, or 1024, probabilities, one of which is associated with 0000000000 and another with 1001011001, and so on. The situation is nearly identical, except for the fact that instead of complex numbers, the probabilities are always positive. Now, suppose I have a probabilistic computer that simulates the flipping of ten coins. It manipulates numbers for each of the 1024 possible sequences of heads and tails just like a quantum computer would. So, does it calculate those probabilities in parallel universes? No, obviously not. But clearly, there must be a difference between the two computers.

When you have a list of probabilities representing bits of information, and you change those bits of information—by processing them in a computer, say—then the list of probabilities obviously changes. In general, the new list is a mixed-up version of the old list obtained by multiplying and adding the original numbers together. But here's the thing: if all those numbers are positive, they can never cancel each other out. Multiplying and adding positive numbers always results in positive numbers. Meanwhile, with qubits, the list can change in drastically different ways because adding negative numbers to positive numbers can lead to cancellation. In other words, a classical computation is just a quantum computation restricted to positive numbers that add up to one.

Rather than picturing quantum computers as some exotic new addition to our classical world, let's turn things around. Imagine a world fundamentally quantum, where the large objects we're familiar with are strangely constrained. They can only perform a specific kind of computation without the ability to cancel out possibilities the way full-fledged quantum systems can.

## Deflating the multiverse

Quantum computations happen in this universe, not the multiverse. But the media, always on the hunt for tantalizing stories, grabbed onto this narrative, creating a feedback loop. The more the idea was mentioned, the more ingrained it became in public consciousness. Over time, the concept of many worlds became intertwined with quantum computing in popular discourse, leading to the prevalent yet misconstrued belief that quantum computers work by operating simultaneously across parallel universes.

The portrayal of quantum computing as a magical tool that can solve all problems by computing in multiple universes can lead to misunderstandings and inflated expectations. It's crucial to separate the fascinating yet speculative ideas about the nature of reality from the actual, proven capabilities of quantum computers. While Deutsch was inspired by the MWI and sees quantum computers as evidence for it, the actual operation and utility of quantum computers don't require MWI to be true. In other words, quantum computers work based on the principles of quantum mechanics, and their functionality is independent of the philosophical interpretation of those principles.

# Myth 4: Quantum Computers Communicate Instantaneously

> *"I cannot seriously believe in it because the theory cannot be reconciled with the idea that physics should represent a reality in time and space, free from spooky action at a distance."*
>
> — Albert Einstein

Entanglement is surely the most misunderstood concept in quantum physics, often depicted as a kind of mystical connection enabling instant communication across vast distances. This misunderstanding has led to widespread speculation about quantum computers exploiting this phenomenon to achieve instantaneous data transfer. But like the myths before it, this concept of "spooky action at a distance" as a computational resource strays far from the realities of quantum mechanics and the operational principles of quantum computing.

While the quantum data within quantum computers is indeed *entangled*, we can reframe our understanding such that this should seem inevitable rather than miraculous.

## Where did entanglement come from?

In 1935, Albert Einstein and two colleagues, Boris Podolsky and Nathan Rosen, wrote a paper elucidating the conceptual problem quantum physics posed for our classical notions of space and time. Either quantum physics disobeyed Einstein's theory of relativity, or quantum physics was not a complete theory. Einstein dismissed the former—the infamous "spooky action at a distance"—suggesting that there must be some deeper reality behind the equations of quantum physics. Since the theory didn't specify what these might be, they came to be called "hidden variables."

After a New York Times headline read "Einstein Attacks Quantum Theory," attention was drawn to the feature Einstein, Podolsky, and Rosen identified. Erwin Schrödinger was the first to name it and its own English translation, calling it *entanglement*. Of it, he said entanglement is "the characteristic trait of quantum mechanics, the one that enforces its entire departure from classical lines of thought." Then, he left physics to become an Irish biologist. Besides some wordy and pedantic public debates between Einstein and Niels Bohr, not much more was said about entanglement for decades. War and the shift from science to engineering in quantum physics produced the "shut up and calculate" generation, which frowned upon what they saw as fruitless philosophical matters. Of course, there are always a brave few.

One of the brave souls was John S. Bell. In 1964, he proposed an experiment that could rule out exactly the kind of hidden variables that the now-late Einstein hoped for. His proposed experiment—and many refinements that came later—are known as *Bell experiments*. The quantities these experiments measure are compared to *Bell inequalities*. Entanglement is necessary to violate the inequalities, which in turn rule out local hidden variables. John Clauser and his student, Stuart Freedman, performed the first real Bell experiment eight years later. This could have ended the debate, but the hope for hidden variables was strong. People began looking for so-called "loopholes" that might leave room for Einstein's desires. Alain Aspect and Anton Zeilinger followed with their own experiments to close the loopholes a decade later. In doing so, they paved the way for extremely precise control of entangled quantum states, which ushered in a new era of quantum technology.

## Explainer-level nonsense

The gist of any quantum entanglement story is that it arises when particles interact and create a "link," which we call entanglement. Importantly, entanglement remains no matter how far apart they might be. The state of each individual particle is not well defined, but their joint (entangled) state is. Thus, the two particles must be considered as a single entity spread across a potentially vast distance. If you believe that story, I agree it's mystical, just as the internet told you.

The mechanics of what is going on when actually creating entangled states of quantum things seem to corroborate this story, though. We've been creating entangled states of photons, for example, this way for decades. A single high-energy photon enters a special transparent material that converts it into two lower-energy photons that fly off in different directions. Because of an ambiguity in which photon has which property, quantum physics says they are entangled. Numerous experiments have verified the predictions of measuring the two entangled photons, with the latest separations being tens of kilometers. In fact, each new experiment has accompanying press releases staking their claim on the current quantum entanglement distance record.

The fact that physical properties, like polarization in the case of photons, are correlated in these experiments makes it feel like a real link has been created. But that's not the only way to create entanglement.

## Cutting the link

Quantum mechanics makes accurate predictions in the context of entangled systems. It doesn't actually contain or suggest the model of entanglement as a physical connection between two distant objects, though. That model is wrong. To see why, consider that entanglement can be created between two particles without ever having them interact. They can be so far away from each other that not even light signals can reach one another, meaning nothing physical could have mediated the "link."

First, imagine two atoms separated by a large distance and both in an excited state. When either atom decays, it releases a photon that is detected at a central station midway between the two distant atoms. Since the detector cannot distinguish which of the two atoms decayed, the state of the pair becomes correlated. Quantum mechanics dictates they must be entangled, but it says nothing about a physical link.

The atoms never interacted or exchanged any information. In fact, they couldn't have—the only signal that something had happened made it halfway between them when the entanglement was created. Are we to believe a physical link of double that size was immediately generated? I hope not. So, what's going on then?

## Classical entanglement

Imagine if, instead of atoms, there were two distant boxes, each with a ball in it. The ball might be removed and sent to you from either box. At some median location, you receive a ball in the post. Immediately, the boxes become correlated—one is empty, and the other is not—because you don't know which box the ball came from. While the boxes, still separated by a great distance, instantly formed this connection, it is simply your ignorance and future expectations about what might be revealed that define the correlation. There is certainly no mystical "link" that physically manifested between the boxes the moment the post arrived, and the same is true for atoms. The point here is that, like correlated bits, entanglement is simply correlated *qubits*.

Now, of course, there must be some difference between the very classical "balls in boxes" situation and the quantum atoms. In the classical world, correlated events cause one another or can be traced back to a common cause that could have determined the outcome. An infamous example is the fact that cities with more police have more crime. Neither causes the other, though. The confounding factor is city size—bigger cities have *both* more police and more criminals simply because they have more people. There is always something that explains correlations in classical information. In the case of the balls, in principle, someone could know the whole situation—which box was empty and where each ball was. That's just not possible with atoms and photons.

When we attempt to "explain" quantum correlations, we naively and unavoidably constrain ourselves to stories phrased in classical information. These are essentially hidden variables, which we ought to know won't do. You can explain quantum entanglement, but it must be phrased in quantum information. Demanding a classical explanation of entanglement is like demanding the behavior of rabbits be explained in terms of apples.

## Technobabble

The point of mathematics is simplifying things that would require otherwise long-winded and complicated sentences. So, we replace the things we are talking about with symbols and numbers. If classical bits are unknown, we write them as a list of probabilities ($p_1$, $p_2$, $p_3$, …). In the case of the two boxes, our ignorance of the contents of each box is a probabilistic bit, as introduced in the previous chapter. The first box is associated with a pair of probabilities ($p_1$, $p_2$). Again, this is just a way more succinct way than writing (the probability that this box has a ball, the probability that this box has no ball). Between the two boxes, there are four possible situations, which would have a list of numbers like ($q_1$, $q_2$, $q_3$, $q_4$).

Now, here's the important point: if the list of four probabilities for the pair of them can't be equally described as two separate lists of two numbers for each of them, then the information they share must be correlated. Mathematically, you can take this as the definition of correlation. For example, (0, 0.5, 0.5, 0) represents the situation when one ball is received at the central location. There is zero chance both are empty and zero chance both have a ball. We are certain one is empty, and one has a ball—we just don't know which is which. Since we don't know what box it came from, either each box is empty or contains a ball with 0.5 probability. Each box alone has the same probability pair (0.5, 0.5), but these individual lists don't capture the complete situation—a bigger list is always needed to capture the correlations.

You now know what regular old classical correlation is. Luckily, entanglement is not much different. Let's recall the definition of a qubits. Instead of two positive numbers that add up to one, a qubit is represented by two numbers (which could be negative) that add up to one after you square them. For example, (0.6, −0.8) represents a qubit. Clearly, these are neither positive nor do they add to one. But if you square each of them, you get (0.36, 0.64), which adds to one. When you square the numbers in the qubit list, it tells you the probability of each possible answer to the question the qubit represents. If one atom is described by a qubit (0.6, −0.8), then we would find it excited with a probability of 0.36 and decayed with a probability of 0.64.

For the pair of atoms, the two-qubit state has four numbers. For example, (0, 0.6, −0.8, 0) tells us that only one atom will be found in the excited state, but with unequal probabilities. If the list representing the pair of atoms can't be equally represented by two smaller lists for each atom individually, they are correlated. But since they are qubits instead of bits, we give such a list a new name: entanglement. That's it. In quantum information, entanglement is correlated qubits.

## Interference

Lists of probabilities change by multiplying and adding up the individual numbers to create new ones. As pointed out in the last chapter, multiplying or adding positive numbers can only produce more positive numbers. Whereas, with qubits, the list can change in drastically different ways because adding negative numbers to positive numbers can lead to cancellation. Borrowing terminology from wave mechanics, this is often referred to as *interference*, where two waves cancel when the crest of one meets the trough of another.

Quantum computers perform calculations in far fewer steps than classical computers by using interference—

choreographing the cancellation of unwanted numbers in qubits of information. While the computer doesn't use entanglement as some physical fuel, we can show that without it, the computations it performs can be easily simulated with classical digital computers. That is, a quantum computer, made of atoms and photons, for example, that never realizes entanglement is no less "quantum" than anything else but also no more powerful than a digital computer. In some sense, though, this is not surprising. After all, a digital computer that never produces correlated bits would be extremely useless—no more powerful than flipping a bunch of fair coins.

## Beam me up

Correlated qubits are necessary for quantum computation and can be seen as a resource for primitive information-processing tasks. The whimsical names of these tasks don't help our myth-busting endeavor, however.

Take, for example, quantum teleportation. Entanglement is necessary to "teleport" quantum information between locations using only classical information. While this doesn't mean teleportation in the sci-fi sense of instantly transporting matter, it does involve the transfer of quantum information in a way that's not possible without quantum entanglement. When thinking about entanglement as a physical connection, things like teleportation do indeed sound like science fiction. However, quantum teleportation is just shifting the location where information is stored in an efficient way. If I were narrating the protocol, I might say the following.

*"Two qubits are correlated. I take one and compare it to a third. Now, two of the three qubits have been read. Having learned something about that comparison, I manipulate the unread qubit so that it's described in the same state as the third one before reading it."*

It's not that you are meant to follow the logic there—the protocol itself is not trivial. However, compare this to an explanation from Popular Mechanics (A. Thompson, March 16, 2017).

*"If we take two particles, entangle them, and send one to the moon, then we can use that property of entanglement to teleport something between them. If we have an object we want to teleport, all we have to do is include that object in the entanglement… After that, it's just a matter of making an observation of the object you want to teleport, which sends that information to the other entangled particle on the moon. Just like that, your object is teleported, assuming you have enough raw material on the other side."*

Thinking about quantum information as physically corresponding to classical objects quickly descends into magical thinking. Not only does it not help explain the concepts, but it further mystifies quantum physics and gives it the illusion that supernatural forces are at play.

## Just correlations

Most of what you hear and read about quantum entanglement is the shooting down of attempts to force it into a classical worldview, but framed with headlines like "quantum physicists just proved nature is spooky." Technically, we call the results no-go theorems because they rule out theories that would restore classical objectivity to quantum physics. Classical objectivity is comforting because it provides a reliable and persistent model of the world. It allows us to predict and control our environment with remarkable ease as we cobble together rigid objects to act as simple machines that extend our natural abilities and more complicated ones that have enabled a mostly cooperative global technological society.

We found and exploited regular patterns in the world. But the comfort of objectivity is an illusion—an illusion that has made "physics" synonymous with "objective reality." This is not a problem for classical information since there is a perfect correspondence between bits and easily recognizable binary alternatives in the objective world we have created.

But that's a very narrow view influenced by the success of classical physics and engineering. Quantum physics, with things like entanglement, throws a wrench into this neat classical picture of the world.

Yes, entanglement challenges our deepest desires for a universe of rigid cause-and-effect relationships. However, as noted when debunking Myth 1, unless you "speak" quantum information natively, you won't "understand" entanglement in the mechanical way you understand most other things. But, from a higher vantage point, you can appreciate it. Classical correlations are relationships between bits of information. The easiest high-level description of entanglement is the correlation between quantum bits (qubits) of information. In the same way that correlated bits are ubiquitous and inevitable, so is entanglement. Quantum bits belong to a theory of information where correlation is the norm, not the exception. Sure, to encode entangled qubits faithfully into the world might be an engineering challenge, but the concept is ultimately substrate-independent, living in the abstract realm of information and algorithms—no mysterious links, instantaneous communication, or spooky actions.

# Myth 5: Quantum Computers Will Replace Digital Computers

*"Create the hype, but don't ever believe it."*

— Simon Cowell

We are immersed in a "replacement culture" that assumes new technologies will completely supplant their predecessors. This mindset, which has witnessed the transition from records to CDs to streaming or from landlines to wireless handsets to mobile phones, might naturally lead many to assume that quantum computers will replace digital ones.

It's easy to be caught up in the buzz and excitement surrounding emerging technologies, especially something as ground-breaking as quantum computers. Predictions of them replacing our everyday devices, from laptops to smartphones, permeate popular culture and the media. This is a form of hype that misrepresents the potential of quantum computing.

This chapter will first describe the likely trajectory of quantum technology and detail why conventional computers will always be needed and, in some cases, remain superior to quantum computers.

## GPUs to QPUs

As mentioned in the introductory chapter, a quantum computer is unlikely to be a "computer," as we colloquially understand it, but a special-purpose processing unit—the QPU. The term QPU (quantum processing unit) is analogous to the graphics processing unit (GPU).

The history of the GPU begins in the late 1990s, with Nvidia introducing the GeForce 256 in 1999, often celebrated as the first GPU. Originally, GPUs were designed to accelerate graphics rendering for gaming and professional visualization, offloading these tasks from the CPU (central processing unit) to increase performance and efficiency.

Over the years, GPUs evolved from specialized graphics accelerators into highly parallel, multi-core processors capable of handling a wide range of computing tasks. This transition was facilitated by developing programming models like CUDA (Compute Unified Device Architecture), which allowed developers to use GPUs for tasks beyond graphics display, including scientific research, machine learning, and something to do with cryptocurrencies.

Adopting GPUs for parallel processing tasks significantly accelerated computations in fields requiring intensive mathematical calculations, transforming industries and research. This shift underscored GPUs as indispensable for specific applications rather than general computing needs.

Currently, QPUs are in their nascent stages, with ongoing research focused on overcoming significant challenges such as error rates, qubit coherence times, and scalability. Quantum computing is still largely experimental, with prototypes and small-scale quantum computers being tested in research labs and by a few commercial entities. To guess at the quantum technological future is an exercise in crystal gazing, but here we go…

I predict that the developmental trajectory of QPUs will mirror that of early digital computers and GPUs, transitioning from rudimentary and highly specialized devices to more sophisticated and versatile computing platforms. However, due to the inherent complexity and specialized nature of quantum computations, QPUs will evolve into complementary components of classical computers optimized for specific tasks intractable for CPUs. Much like GPUs, QPUs will eventually be utilized beyond the applications we envision for them today, but they will never replace CPUs or even GPUs.

Before we get into what QPUs *might* do, let's outline what they certainly won't do. But before *that*, we need to talk about what problems are hard and why.

## Computational complexity

Theoretical computer scientists categorize computational problems based on the amount of computational resources (like time and space) required to solve them, essentially dividing them into "easy" and "hard" problems.

Imagine you're trying to crack a safe that uses a pin code. Let's start with a four-digit pin, where each digit can be either a 0 or a 1. For the first digit, you have two options. For the second digit, you also have two options, but now, combining it with the first digit's options, you get four total possibilities (00, 01, 10, 11). By the time you reach the fourth digit, you're looking at sixteen possible combinations (0000, 0001, …, 1111). Clearly, this safe is not very secure.

Now, imagine each digit in the pin can be more than just 0 or 1. Let's say there are *n* options for each digit, and the pin is *d* digits long. The total number of combinations is *n* to the power of *d* (or $n^d$, which is *n* multiplied by itself *d* times). If the number of digits *d* stays fixed, but you increase the number of options *n* for each digit, then the total number of combinations grows *polynomially* because polynomials involve variables raised to fixed powers. For example, sticking with a four-digit pin but increasing the options for each digit from just 0 and 1 to, say, 0 through 9, the safe gets more secure because the total combinations jump from 16 to 10,000 (which is $10^4$).

On the flip side, if you keep *n* fixed (like sticking to 0 through 9 and not adding more) but increase the number of digits *d*, you see an *exponential* growth in combinations because the variable changing is in the exponent. This means for each new digit you add, you're multiplying the total number of combinations by *n*, not just adding *n* more combinations. If the new safe has six digits instead of four, we go from 10,000 combinations to a million!

Cracking locks where the number of digits is fixed but the number of options can grow is "easy" because the number of steps grows polynomially. Such a problem belongs to the class labeled P (for *Polynomial time*). Since problems in the class P are already easy for classical computers, quantum computers won't significantly speed them up. An algorithm is called *efficient* if it runs in polynomial time. So, both classical and quantum algorithms for this problem are efficient.

On the other hand, cracking locks where the number of options is fixed but the number of digits can grow is "hard" because the number of steps grows exponentially. Note that even though the problem is identical to the one considered in the previous paragraph for any given lock, problem *classes* require some flexible notion of *input size*. Since the input is different between the two scenarios, the problems really do occupy different classes. When the number of digits can change, the problem belongs to the class NP (Nondeterministic Polynomial time). This class contains all problems for which a solution can be easily verified. In this case, given the correct combination, it is trivial to test it.

For the more difficult problem of cracking a lock with variable digits, the algorithm of trying every combination is not efficient. In fact, there is no efficient classical algorithm for this problem. It's not expected that an efficient quantum algorithm exists for this problem, either. (Though, if you believed the last myth, you might be forgiven for thinking so.)

In summary, quantum computers aren't useful for cracking pin numbers for two reasons. Either the problem is too easy—in which case a classical computer suffices—or the problem is too hard for any computer. By now, you must be wondering what problems a QPU is actually useful for.

## BQP

Ideally, we'd like to define a class of problems that can be solved with a quantum algorithm in polynomial time. In classical complexity theory, most problem classes are defined relative to a *deterministic* algorithm. But quantum algorithms end when quantum data is read, a random process. So, we need to add probability to the definition.

Bounded-error Quantum Polynomial time (BQP) is the class of problems that can be solved with a quantum algorithm in polynomial time with a probability of error of at most ⅓. (The ⅓ is an arbitrary choice, strictly less than ½ if you were wondering.) This class includes all the problems that are "easy" for a quantum computer. Some problems in this class are assumed to be hard for classical computers, but there is some fine print.

Briefly, there are huge, century-old open questions in the theory of computer science that would net you fame and fortune if you solved them. One such example is, does P =

NP? Another: is NP contained in BQP? We suspect the answer is "no" to both of these questions, but there is currently no mathematical proof. This is why many statements about computational speed-up are couched in dodgy language, and some aren't. We don't *really* know if quantum computers are any different from classical computers, but we highly suspect they are. In this book, for brevity, I won't add all the necessary caveats to statements about complexity as if it were a graduate class in computer science.

As mentioned in the introductory chapter, the Quantum Algorithm Zoo ([quantumalgorithmzoo.org](quantumalgorithmzoo.org)) lists problems known to be in BQP—that is, efficiently solvable with a QPU. It's a long list, but most of the problems are extremely abstract. The most famous is probably factoring, which is solved by Shor's algorithm. But that is better suited for the next myth. Here, I'll describe the "obvious" one.

## Quantum simulation

In both classical and quantum physics, scientists use mathematical models called Hamiltonians to describe how things behave—from the interactions of tiny particles to the workings of complex materials. However, understanding and predicting the behavior of these systems can be incredibly complex, requiring massive amounts of computational power, even for supercomputers. Today, many approximations are used to make the problem tractable, yet at the expense of accuracy.

Given a Hamiltonian model, a detailed simulation of it with a digital algorithm is so inefficient that it's not even conceivable that we will ever have the classical computing power to solve it. However, it has been shown that *some* Hamiltonian model simulations can be efficiently carried out on a quantum computer.

At a high level, the idea is both simple and intuitive. Given the Hamiltonian model, break it up into smaller and smaller chunks until each chunk is an elementary quantum logic instruction. Then, carry those instructions out on a universal quantum computer! Of course, it's harder than it appears to do the "chunking," but it intuitively *feels* easy because we are using quantum physics to mimic other quantum physics.

Why is this important? Well, as Feynman said, "Nature is quantum mechanical, damn it." So, if we want to engineer and control Nature at the finest scales, we need to be able to model and simulate it in much the same way we model and simulate everything from bridges to airplanes before we attempt to build them. Indeed, quantum simulation may help scientists and engineers design new materials with exciting properties, like superconductors that work at room temperature or ultra-strong and lightweight materials. These could revolutionize fields like energy, transportation, and electronics. By simulating complex molecules accurately, quantum simulation could speed up drug discovery and even allow scientists to probe exotic new phenomena in a virtual environment free from terrestrial constraints, furthering our understanding of the universe's fundamental workings.

These are surely lofty goals, and this area of quantum algorithm research is a work in progress. We are still far away from a simple pipeline that takes the desired properties of complex systems on one end and pops out solutions on the other. It may be that every potentially revolutionary application is one of the "worst case" scenario problems that even a quantum computer can't efficiently solve. However, that doesn't necessarily render quantum computers useless because two algorithms can be *technically* inefficient while one is much faster than the other.

## Brief aside on solving "hard" problems

We try to solve "hard" problems all the time. Just because a problem doesn't omit an efficient algorithm doesn't mean we can't solve it—we simply need to evaluate how much time and energy we are willing to put into solving it. Historically, algorithm breakthroughs have occasionally turned the tide, making previously intractable problems manageable. A classic example of this is the development of the Fast Fourier

Transform (FFT) and its profound impact on signal analysis, which can serve as a bridge to understanding the potential of quantum algorithms that don't omit exponential speed-ups.

The FFT is an algorithm for computing the Discrete Fourier Transform (DFT) and its inverse. The DFT is used across various fields of science and engineering, from signal processing to solving partial differential equations. Before the introduction of the FFT, computing the DFT was a computationally expensive task, requiring $n^2$ operations for $n$ data points. The FFT reduced this complexity to $n \cdot log(n)$ operations. While this appears to be a modest square root speed-up, it enabled the practical processing of large datasets.

Imagine a communications system such as Wi-Fi using 1,000 subcarriers in its signal. The difference between using the DFT and the FFT in encoding and decoding that signal would be 1,000-fold. That's the difference between one second and fifteen minutes or one day and nearly three years. While it doesn't provide the highly sought-after exponential advantage, the FFT exemplifies how a clever algorithm can change the landscape of computational problem-solving. It allowed for real-time signal processing and practical image compression, among many tasks that were previously prohibitive.

While the hype around quantum computing speed-ups often invokes the term "exponential," we also suspect that QPU can provide square-root speed-ups generically. Again, while this sounds modest, keep in mind the FFT and all that it has done for society when you say that!

## Grover's search algorithm

Imagine you have a massive, unsorted database—think of a disorganized phone book with millions of entries. You need to find a specific person's number, but you don't know where to start. With a classical computer, machine, or just yourself, you'd have to check the entries individually. On average, you'd have to search through about half the database before finding the right one. In the worst case, you'd end up checking the entire phone book!

While we are unlikely ever to find a scrabbled-up phone book, many problems can be recast as essentially the same thing. Finding the correct combination for a pin number is a perfect example. Thus, solving this problem faster is practically relevant, which is where quantum fits in. Grover's search algorithm is a quantum algorithm that speeds up the process of finding a marked item in an unsorted database. How does it do that? First, as a fun exercise, let's give the popular version that utilizes the myths already discussed.

Quantum-powered search. Instead of checking items one by one, a quantum computer can use superposition to examine all database entries at the same time. It's like looking at all the pages of the phone book simultaneously. Okay, now you know that's not really how it works. So, what's a better explanation than simply "complex math?"

Recall from previous chapters that qubits are long lists of numbers that can be both positive and negative. Steps in an algorithm amount to multiplying and adding these numbers together to get a new list. When positive and negative numbers combine, they tend to produce smaller numbers, while like-signed numbers reinforce each other to produce larger numbers. Grover's algorithm works by first assuming some "oracle" exists that multiples the number at the location of the marked item by −1. So, all the numbers are the same, except for one, but it's "hidden"—the algorithm can't know which it is. The next step is *diffusion*, which spreads the numbers around like a wave dissipating. Equal numbers don't affect each other, which we would expect from the symmetry of the situation. However, each number shares an asymmetric relationship with the marked number. The effect is that every number has a bit of its value shaved off and given to the marked item's number.

By repeating the above process, the number corresponding to the marked item becomes larger and larger while every other number approaches zero. Think of it as the right entry becoming brighter while all others fade. The algorithm cleverly manipulates the quantum information so that the marked entry becomes more likely to be found when the data is finally read.

The steps in Grover's algorithm use the intuition of how waves behave and where interference will occur, which is completely different from how digital algorithms are designed. This illustrates both the difficulty and potential of quantum algorithms. They require new insights to develop, which means many problems we haven't even thought about could be amenable to them.

While Grover's algorithm gives *merely* a square root speed-up over exhaustively searching databases, recall the FFT and its implications.

# Myth 6: Quantum Computers Will Break the Internet

*"Amateurs hack systems, professionals hack people."*

— Bruce Schneier

Y2Q, Q-Day, Quantum Leap Day, Q-Break, the Quantum Apocalypse—whatever you want to call it—the day quantum computers are destined to shatter the internet, decrypting the world's secrets as easily as a child unwraps their birthday gifts. In anything but hushed tones, we've been warned of a digital apocalypse, where privacy crumbles and chaos reigns, all with the flick of a quantum switch.

While it is true that a quantum algorithm exists that can decrypt most messages sent over the internet, the reality of using it is a different story altogether. Of course, like most stories, we need to start at the beginning.

## A brief history of internet encryption

The internet's precursor, ARPANET, was developed as a project by the Advanced Research Projects Agency (ARPA) of the U.S. Department of Defense in the late 1960s. Encryption at this time was primarily a concern for military communications. Crudely speaking, there are only two types of people in this context: us and them. Every physical device can carry the same secret used to encrypt messages. However, in a non-military context, every *pair* of people in the network needs to share a separate secret. Every new person added to the network would have to find some way of transmitting a unique new secret to every person already in the network. This was a seemingly infeasible proposition until the mid-1970s.

In 1976, Whitfield Diffie and Martin Hellman introduced public key cryptography, fundamentally changing the way encryption could be used over networks. This method allowed two parties to create a shared secret over an insecure network channel. A year later, the RSA (named for Ron Rivest, Adi Shamir, and Leonard Adleman) algorithm was published, providing a practical method for secure data transmission. This algorithm become the foundation for secure communications on the internet, which is simply the amalgamation of multiple networks with standardized communication protocols.

In the 1990s, as the internet became more commercialized and accessible to the public, the need for secure transactions became apparent. The introduction of the Secure Sockets Layer (SSL) protocol by Netscape in 1994 was a critical step in enabling secure online communications and transactions. By the 2000s, with the increasing prevalence of cyber threats, encryption became indispensable. Technologies like Virtual Private Networks (VPNs), end-to-end encryption in messaging apps, and secure browsing protocols for websites become standard practices. Today, it is simply assumed that your interactions on the internet are private. But why and how?

## To crack a code

To understand the challenge of breaking internet encryption, especially using RSA, let's simplify the layers and protocols into a more digestible scenario. Imagine you're shopping online at your favorite bookstore and decide to purchase a new book from your favorite author (*winky face emoji*). At checkout, you're prompted to enter your credit card details. This is where RSA encryption steps in to protect your information.

When you hit "submit" on your credit card details, your browser uses the bookstore's public key to encrypt this information. The process, devoid of the mathematical symbols you probably don't want to see while enjoying a quick read, can be likened to locking your credit card details in a box. The bookstore has the only key (the private key) that can unlock the box. Even though the locked box travels across the vast and insecure network of the internet, only the bookstore can open it upon receipt, ensuring that your credit card information remains confidential. However, by inspecting the lock carefully enough, you could deduce what the (private) key looks like, make one, and unlock the box. The question is: how long would this "breaking" of the lock take?

The security of this transaction relies heavily on the RSA algorithm's use of large prime numbers to generate both the public and the private keys. The public key is openly shared, while the private key remains a closely guarded secret of the bookstore. The strength of RSA lies in the fact that, with current computing power, it is virtually impossible to deduce the private key from the public key due to the difficulty of factoring the product of the two large primes used in their creation.

Consider the following number:

25195908475657893494027183240048398571429282126
20403202777713783604366202070759555626401852588
07844069182906412495150821892985591491761845028

08489120072844992687392807287776735971418347270
2618963750149718246911650776133798590957000973 3
0459748808428401797429100642458691817195118746 1
2151517265463228221686998754918242243363725908 5
1418654620435767984233871847744479207399342365 8
4823824281198163815010674810451660377306056201 6
1967625613384414360383390441495263443219011465 7
5444541784240209246165157233507787077498171257 7
2467962926386356373289912154831438167899885040 4
4536402352738195137863656439121201039712282212 0
720357.

This number (2048 bits long if written in binary) is the product of just two prime numbers (numbers without factors). It's not difficult to create such a number—simply take two large prime numbers and multiply them together. Your calculator could perform this feat almost instantly. The hard part is doing the reverse, factoring the larger number into its smaller components. I didn't do that for the number above. That number is RSA-2048. At the time of writing, only RSA Security LLC, the company that published it, knows which two prime numbers were multiplied together to produce it. In fact, if you can "crack" it, they'll give you $200,000.

The largest RSA number factored as part of this challenge was 768 bits long. It took the equivalent of a single computer

2,000 years to solve. In short, if someone could solve this so-called *Integer Factorization Problem* efficiently, they'd "break" the internet and blow some minds because of how difficult it appears to be. Shor's algorithm, which has been mentioned several times, does exactly what internet security and cryptography experts thought was impossible—it efficiently solves the integer factorization problem.

## Enter the quantum

Peter Shor introduced his factoring algorithm in 1994 at the 35th Annual Symposium on Foundations of Computer Science. The reaction was a swift mixture of skepticism, excitement, and concern.

As with any sudden and major breakthrough, many were justifiably skeptical. Quantum computing was still very theoretical, with only a handful of researchers working in this fringe field. As other researchers dug into the details, they realized its theoretical significance. It wasn't just its potential to break encryption but the fact that it represented one of the first convincing examples of a quantum computer outperforming classical ones by a vast margin.

More than just a theoretical curiosity, Shor's algorithm galvanized research into quantum computing. It provided a tangible goal—if a quantum computer could be built, it would have immediate and dramatic practical implications. This was of obvious concern to cryptographic security experts who were still primarily deployed in a military context.

Nowadays, we routinely teach Shor's algorithm to undergraduate students. The lesson typically appears near the end of a semester-long course on quantum computing, which assumes the students have been introduced to far more technical and mathematical background than provided in this book. However, the gist can still provide insights.

Much like Grover's search algorithm, which solves the search problem in a way completely different from its digital counterpart, Shor's algorithm takes advantage of the way quantum data represents complex numbers. The best-known classical algorithm for factoring is called the *General Number Field Sieve.* In many ways, it's more complicated than Shor's algorithm. Indeed, quantum computers don't appear to be useful aids in any of the steps in the best classical algorithm. Shor's algorithm takes a more direct route to solving the problem—one not taken in classical algorithms because it's far less effective.

Shor realized that factoring can be broken down into steps, and most of the steps are easy (even for digital computers), but one step is hard. The hard step is finding the pattern in a periodic function. While this is not *conceptually* hard—in fact, it's the same problem the DFT and FFT discussed above are applied to—it is *computationally* hard for digital computers.

Recall that the list of numbers carried by quantum data is exponentially long in the number of qubits—that's $2^n$ complex numbers for $n$ qubits. The FFT approximates the DFT but would still require more than $2^n$ steps to calculate it for this list of numbers. However, Shor and others showed that the DFT can be applied to the quantum data with a quantum computer using only $n^2$ steps (quantum logic operations), an exponential speed-up.

As a quick aside, note that Shor's algorithm is also not deterministic, underscoring again the critical differences in how we approach digital and quantum algorithm design.

## How far away is it?

As you know, quantum computers today are not powerful enough to apply Shor's algorithm to anything but small toy examples that grade-schoolers could have factored. So, when will they be big enough to break the internet?

In 2021, Craig Gidney and Martin Eker estimated that Shor's algorithm could crack RSA-2048 with a 20,000-qubit quantum computer. This assumes those qubits work more or less perfectly. However, even if they don't, we can rely on redundancy and error correction to get us there. Many poorly performing components can come together to act as a better-performing "virtual" component. Virtual qubits are also called *logical* qubits, whereas the raw physical systems (like atoms and photons) are called *physical* qubits.

Research in error correction is constantly improving, moving toward higher tolerance of errors and lower overheads. Using the best-known techniques and projected error rates, Gidney and Eker suggest that 20 million *physical* qubits would be required. That sounds like a lot at a time when even glossy, hyped-up press releases claim qubit numbers in the hundreds.

Gordon Moore, co-founder of Intel, made an astute observation in 1965: the number of transistors packed into an integrated circuit seemed to double roughly every two years. This observation, later termed *Moore's Law*, became an informal but remarkably accurate predictor driving exponential growth in computing power over decades. The underlying principle is a feedback loop—as computing power increases, chip manufacturing processes improve, enabling even more transistors to be crammed into smaller spaces, leading to a further boost in computational ability.

While quantum computing is still in its early stages, there are signs it may follow a similar exponential growth trajectory—a *quantum* Moore's Law, as it were. If we start with roughly a hundred qubits today, and the number of qubits doubles every two years, it suggests that in about 34 years (around 2058), quantum computers could become powerful enough to crack even the most robust RSA encryption standards. Researchers like Jaime Sevilla and Jess Riedel support this timeline, publishing a report in late 2020 that claimed a 90% confidence of RSA-2048 being factored before 2060.

## Store now, decrypt later

While factoring large numbers is the theoretical target, the practical implications are worth considering today. Take, for example, the online word processor used to write this very text, which employs secure communication using a 256-bit public key. That's much weaker than a 2048-bit key, suggesting it would be vulnerable sooner than the 2058 estimate.

Imagine a nefarious entity collecting and storing encrypted data today. Even without the ability to decrypt it immediately, they could wait until sufficiently powerful quantum computers exist, say around 2058, and then easily decipher the stored secrets. Why does that not bother me?

Well, I don't have 34-year secrets. But government agencies likely do. Secrecy of data has a shelf life, and quantum computers render the immediacy of that more palpable.

While the constant hype about quantum computers can desensitize us from such threats, governments and major corporations have weighed the odds. In late 2023, US security agencies published a factsheet on the impacts of quantum computing, urging all organizations to begin early planning for Q-Day—though, of course, they didn't use that term.

The goal of so-called *post-quantum cryptography* is to develop crypto-systems that are secure against quantum computers running Shor's algorithm and, hopefully, any other quantum algorithm. Ideally, the system would also work with existing communications protocols and networks.

The National Institute of Standards and Technology (NIST) is an agency of the United States government with a mission spanning various domains, including within the realm of information technology and cybersecurity. NIST's involvement in cryptography dates back several decades, with milestones such as developing the Data Encryption Standard (DES) in the 1970s and the Advanced Encryption Standard (AES) in the early 2000s. It's unsurprising then that NIST has been proactive in preparing for the transition to post-quantum cryptography. It launched the Post-Quantum Cryptography Standardization Project in 2016 with a call for algorithms that could withstand attacks from quantum computers.

Since then, NIST has gathered proposals for quantum-resistant cryptographic algorithms from around the world and selected candidates for further analysis and testing by hosting public workshops and soliciting feedback to ensure they meet various criteria. As of the latest update, NIST has announced its intention to publish the new post-quantum cryptographic standards in 2024. These standards will include specifications for quantum-resistant algorithms that have been rigorously tested and evaluated. Major internet companies like Google and Apple have already announced the migration of some services to include additional security layers with quantum-resistant techniques. So, the internet may be safe after all! (Don't worry, your Bitcoins will likely be safe as well.)

## The quantum future

Amidst concerns of a quantum apocalypse, it's vital to differentiate between the messengers of doom—often amplified by tech journalism—and the voices of the quantum research community. The latter seeks not to incite fear but to illuminate the path toward understanding the ultimate limits of our ability to know and control the natural world. In the viral vortex of discussions surrounding quantum computing's potential to disrupt current cryptographic standards, a beacon of hope shines through: Quantum Key Distribution (QKD).

In a typical QKD scenario, two people wishing to communicate securely use a *quantum* channel to exchange qubits that will be used to encode bits of a secret key. The quantum properties of these qubits ensure that any attempt by an eavesdropper to intercept and read the key would irreversibly alter their state, thus revealing the intrusion. The fact that reading qubits destroys the quantum information is not a bug but a feature!

Once a secret key is successfully exchanged and its integrity verified, it can be used to encrypt and decrypt classical messages using conventional cryptographic techniques. Thus, the security of communication is guaranteed not by the hardness of computational problems but by the inviolable principles of quantum physics.

QKD heralds a future where communications can be secured against any adversary, regardless of their computational power. This quantum-safe key distribution method could revolutionize how we protect sensitive information, rendering it immune to the threats posed by quantum computing advancements.

Since we don't have quantum channels lying around, implementing QKD on a global scale faces technological and infrastructural challenges. Specialized hardware will be required to mitigate the difficulty of transmitting qubits over long distances without degradation. Research and development in quantum repeaters and satellite-based QKD are among the avenues being explored to overcome these obstacles.

While Y2Q might get clicks, quantum computing will eventually bring about the opposite of disaster—a global quantum internet providing an additional uncrackable layer of security.

# Myth 7: Quantum Computing Is Impossible

> *"You insist that there is something a machine cannot do. If you tell me precisely what it is a machine cannot do, then I can always make a machine which will do just that."*
>
> — John von Neumann

Quantum computing is not without its detractors. Their arguments come in two flavors. The first style of argument against quantum computing is usually simplified as quantum systems are too complex to control at the level necessary for reliable computation. Recall the exponentially long list of numbers needed to specify a qubit. How could we possibly keep track of and control all those? This is not only defeatist but also naive. Complexity often manifests from simple rules, which we seek to manipulate to build up large quantum devices.

The other argument against quantum computing is that noise and errors will be so unavoidable that quantum computations will break down before answers can be reached. Naysayers need only point to existing quantum computers to illustrate their claim, not to mention the hype, empty promises, and failed predictions of "within the next five years."

Let's not dive straight in but take a step back.

## What really is a quantum computer?

What is called a quantum computer today seems to be a matter of taste. Consider the fact that my smartphone works by flipping bits encoded in transistors that are only a few dozen atoms thick. There are about a billion of them inside my phone, which could not be made without understanding quantum physics. But my smartphone is not a quantum computer—it's a digital computer. This is because it encodes bits—digital, classical data—and not qubits, the building blocks of quantum data.

But wait. It's not that simple. I can indeed encode qubits quite faithfully with the transistors in my smartphone. My smartphone could quite comfortably emulate a perfect quantum computer, provided it had less than 30 qubits or so. When I hijack all the computing resources in my phone to do this, it really is a 30-qubit quantum computer.

But you could argue that in that scenario, the qubits aren't *physical*. When I googled "what is the most powerful quantum computer," the top hit was a press release from a company I won't name claiming that their quantum computer was the most powerful. The device encoded one qubit of information onto each of six individual atoms isolated in a specialized ion trap sitting on an integrated chip—a marvel of modern engineering, no doubt. In such a device, you can make a correspondence between the information you want to encode and the information needed to describe each atom. Thus, qubits are physical.

But... not quite. Once we have large, reliable quantum computers, the qubits will not be encoded onto individual atoms or any other quantum degrees of freedom, for that matter. Error-correcting codes will be used to encode the qubits virtually to protect them from physical errors. This is part of the reason why simply counting qubits in a device doesn't reveal its true utility and why one company can claim their device is "more powerful" than competing devices, which might have ten times as many qubits. I could facetiously pick up a pile of dirt and say it is a quantum computer since each dirt atom encodes quantum data. The problem is, I wouldn't be able to reliably encode it, read it, and certainly not process it. My 30-qubit smartphone, however, does reliably encode quantum data. In fact, by some accepted measures, it is the most powerful quantum computer!

Many companies quote the "quantum volume" of their devices. At the time of writing, a solid 10-qubit device would have a quantum volume of 1024, for example. The quantum volume measures the difficulty a digital computer would face in emulating a given quantum device. It is 2 to the power of the number of qubits (provided that the same number of layered instructions could be carried out without error). Ten qubits that can carry out ten complex instructions without error thus have a quantum volume of $2^{10}$ = 1024. My smartphone can carry out instructions perfectly forever, and so it has a quantum volume of $2^{30}$ or about a billion. (This is not surprising because it has 8 billion bytes of RAM, and it takes about 8 bytes to represent a qubit of data.)

Alright, I admit I'm being pedantic. The real reason smartphones aren't called quantum computers is because they are limited to 30 qubits, and there is no way the technology can scale significantly beyond that. Meanwhile, there is hope for quantum hardware companies that the devices they are building can scale up the number of qubits and the errors down indefinitely.

Maybe they shouldn't call their devices quantum computers today, though—at least not until they outperform my smartphone. The question then is when will that happen? The nice thing—if you really think this is a problem—is that quantum computing is a divergent technology. Far enough into the future, there will be no ambiguity about whether a computer device is *really* a quantum computer because there will not be enough atoms in the universe to construct smartphone memory capable of encoding the quantum data the new device can carry. So, one solution is to just hold off on our bickering until that time.

## What will it take to get there?

In the early days, quantum computers were built by graduate students in poorly funded physics labs. Ever the industrious types, these physicists constructed prototype devices that were meticulously concocted Rube Goldberg machines as much as they were quantum computers. Exposed cables and duct tape are sufficient to demonstrate the concept but do not feature in any path to scalability. Yet, this exemplifies both the challenge and the hope for quantum computers. Yes, new supporting technologies will need to be developed, but also new supporting technology *will be* developed because that's what a bunch of humans collaborating on solving problems tend to do.

Let's consider an example.

Superconducting qubits are so named because they use superconducting circuits cooled to extremely low temperatures. This is done to maintain their superconducting properties and limit thermal (heat) fluctuations. Each qubit is an "artificial atom" made using the same techniques as conventional microelectronics fabrication. Why use existing infrastructure to build something radically new? Well, you work with what you got. However, this aspect of building quantum computers is not a bottleneck. Indeed, specialized superconducting chip foundries are cropping up to address the specialized needs of quantum technology.

Once constructed, maintaining near-absolute-zero temperatures for many qubits requires advanced and scalable cryogenic technology. These needs have pushed existing providers to innovate, creating bespoke solutions specifically for quantum computer chips. Early examples of superconducting qubits had all of the electronics required to control and read them sitting outside the "fridge." This was borne out of the necessity of experimentation to separate systems and because conventional electronics weren't designed for cryogenic temperatures. Today, researchers are designing ways to bring the control and readout logic to the level of the qubits, closing yet another gap in full integration. And so on it will go.

Each company involved in building quantum computers has some vision of how their technology will scale, but look carefully at their roadmaps, and you'll notice quite a few gaps where "miracles" need to occur. But such breakthroughs are inevitable. History shows us that innovation often comes from unexpected places, such as the discovery that graphene could be produced with a pencil and sticky tape. These moments of serendipity, combined with relentless pursuit and ingenuity, are what will bridge the gaps on the roadmap to scalable quantum computing.

Bootstrapping—the iterative process where each advancement builds upon the last—will play a crucial role in quantum computing's development. Just as tomorrow's CPUs are designed using today's CPUs, quantum computing will likely follow a similar evolutionary trajectory, with each generation of quantum computers facilitating the design of its successor. Take a careful look at the design of any complex system, and you will quickly convince yourself that no one person can comprehend all of it. Such systems are built up over years and decades, the collective effort of many individuals collaborating with the very technology they are trying to scale.

## The final coin analogy

Find a coin. Flip it. Did you get heads? Flip it again. Heads. Again. Tails. Again, again, again… HHTHHTTTHHTHHTHHTTHT. Is that what you got? No, of course, you didn't. That feels obvious. But why?

There are about 1 million different combinations of heads and tails in a sequence of 20 coin flips. The chance that we would get the same string of H's and T's is 1 in a million. You might as well play the lottery if you feel that lucky. (You're not that lucky, by the way—don't waste your money.)

Now imagine 100 coin flips or maybe a nice round number like 266. With just 266 coin flips, the number of possible sequences of heads and tails is just larger than the number of atoms in the entire universe. Written in plain English the number is 118 quinvigintillion 571 quattuorvigintillion 99 trevigintillion 379 duovigintillion 11 unvigintillion 784 vigintillion 113 novemdecillion 736 octodecillion 688 septendecillion 648 sexdecillion 896 quindecillion 417 quattuordecillion 641 tredecillion 748 duodecillion 464 undecillion 297 decillion 615 nonillion 937 octillion 576 septillion 404 sextillion 566 quintillion 24 quadrillion 103 trillion 44 billion 751 million 294 thousand 464.

So, obviously, we can't write them all down. What about if we just tried to count them one by one, one each second? We couldn't do it alone, but what if everyone on Earth helped us? Let's round up and say there are 10 billion of us. That wouldn't do it. What if each of those 10 billion people had a computer that could count 10 billion sequences per second instead? Still no. OK, let's say, for the sake of argument, that there were 10 billion other planets like Earth in the Milky Way, and we got all 10 billion people on each of the 10 billion planets to count 10 billion sequences per second. What? Still no? Alright, fine. What if there were 10 billion galaxies, each with these 10 billion planets? Not yet? Oh, my.

Even if there were 10 billion universes, each of which had 10 billion galaxies, which in turn had 10 billion habitable planets, which happened to have 10 billion people, all of which had 10 billion computers, which could count 10 billion sequences per second, it would still take 100 times the age of all those universes to count the number of possible sequences in just 266 coin flips. If that's not the most mind-blowing thing you've read today…

Why am I telling you all this? The point I want to get across is that humanity's knack for pattern finding has given us the false impression that life, nature, or the universe is simple. It's not. It's actually really complicated. But like a drunk looking for their keys under the lamp post, we only see the simple things because that's all we can process. The simple things, however, are the exception, not the rule.

Suppose I give you a problem: simulate the outcome of 266 coin tosses. Do you think you could solve it? Maybe you are thinking, *well, you just told me that I couldn't even hope to write down all the possibilities. How could I possibly choose from one of them?* Fair. But, then again, you have the coin and 10 minutes to spare. As you solve the problem, you might realize that you are, in fact, a computer. You took an input, you are performing the steps in an algorithm, and you will soon produce an output. You've solved the problem.

A problem you definitely could not solve is to simulate 266 coin tosses if the outcome of each toss depended on the outcome of the previous tosses in an arbitrary way as if the coin had a massive memory bank. In that case, you'd have to keep track of all the possibilities, which we just decided was impossible. Well, not impossible, just really, really, *really* time-consuming. But all the ways that one toss could depend on previous tosses is yet even more difficult to count—in fact, it's uncountable. One situation where it is not difficult is the one most familiar to us—when each coin toss is completely independent of all previous and future tosses. This seems like the only obvious situation because it is the only one we are familiar with. But we are only familiar with it because it is one we know how to solve.

Life's generally complicated, but not so if we stay on the narrow paths of simplicity. Computers, deep down in their guts, are making sequences that look like those of coin flips. Computers work by flipping transistors on and off. But your computer will never produce every possible sequence of bits. It stays on the simple path or crashes. There is nothing innately special about your computer that forces it to do this. We never would have built computers that couldn't solve problems quickly. So computers only work at solving problems that we find can be solved because we are at the steering wheel, forcing them to solve problems that appear effortless.

In quantum computing, it is no different. It can be, in general, very complicated. But we look for problems that are solvable, like flipping quantum coins. We are quantum drunks under the lamp post—we are only looking at stuff that we can shine photons on. A quantum computer will not be an all-powerful device that solves all possible problems by controlling more parameters than there are particles in the universe. It will only solve the problems we designed it to solve because those are the problems that can be solved with limited resources.

We don't have to track and keep under control all the details of quantum things, just as your digital computer does not need to track all its possible configurations. So next time someone tells you that quantum computing is complicated because there are so many possibilities involved, remind them that all of nature is complicated—the success of science is finding the patches of simplicity. In quantum computing, we know which path to take. It's still full of debris, and we are smelling flowers and picking the strawberries along the way, so it will take some time—but we'll get there.

# About the author

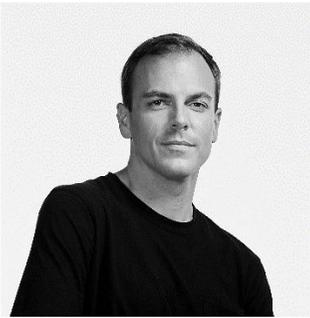

Chris Ferrie is an associate professor at the University of Technology Sydney in Australia, where he researches and lectures on quantum physics, computation, and engineering. He is the author of *Quantum Bullsh\*t: How to Ruin Your Life with Quantum Physics* and with over fifty children's books about science, he is the #1 best-selling science author for kids. You can find Chris online at csferrie.com.